\documentclass[12pt,a4paper]{article}
\usepackage{amsmath}
\usepackage{amsxtra}
    \usepackage{amstext}
    \usepackage{amssymb}
    \usepackage{latexsym}
\usepackage{graphics}

\newcommand{\hepth}[1]{{\tt hep-th/#1}}

\newcommand{\half}{\frac{1}{2}}

\newcommand{\nn}{\nonumber}

\newcommand{\im}[1]{\text{Im}(#1)}

\newcommand{\px}{\partial_{x}}
\newcommand{\pt}{\partial_{t}}

\newcommand{\cb}{{\cal B}}

\newcommand{\p}{\vspace{6pt}\noindent}
\newcommand{\jump}{\vspace{2pt}}

\advance\textheight by \topskip
\makeatletter

\@addtoreset{equation}{section}
\def\section{\@startsection {section}{1}{\z@}{-8.5ex plus -1ex minus
 -.2ex}{3.3ex plus .2ex}{\large\bf}}
\def\subsection{\@startsection{subsection}{2}{\z@}{-3.25ex plus
 -1ex minus -.2ex}{1.5ex plus .2ex}{\bf}}
\def\subsubsection{\@startsection{subsubsection}{3}{\z@}{-3.25ex plus%
 -1ex minus -.2ex}{1.5ex plus .2ex}{\sl}}

\begin{document}
\begin{titlepage}
\vspace*{-2cm}
\begin{flushright}
\end{flushright}

\vspace{0.3cm}

\begin{center}
{\Large {\bf }} \vspace{1cm} {\Large {\bf On
purely transmitting defects
in affine Toda field theory}}\\
\vspace{1cm} {\large  E.\ Corrigan\footnote{\noindent E-mail: {\tt
ec9@york.ac.uk}} and
C.\ Zambon\footnote{\noindent E-mail: {\tt cristina.zambon@ptm.u-cergy.fr}} \\
\vspace{0.3cm}
{${}^a$}{\em Department of Mathematics \\ University of York\\
York YO10 5DD, U.K.} \\
\vspace{0.3cm}
{${}^b$}{\em Laboratoire de Physique Th\'{e}orique et Mod\'{e}lisation \\
Universit\'{e} de Cergy-Pontoise (CNRS UMR 8089), Saint-Martin 2\vspace{.1cm}\\
2 avenue Adolphe Chauvin,
95302 Cergy-Pontoise Cedex, France}}\\

\vspace{1cm} {\bf{ABSTRACT}}
\end{center}

Affine Toda field theories with  a purely transmitting integrable defect are considered
and the model based on $a_2$ is analysed in
detail. After providing a complete characterization of the problem
in a classical framework, a suitable quantum transmission matrix,
able to describe the interaction between an integrable defect and
solitons, is found. Two independent paths are taken
to reach the result. One is an investigation
of the triangle equations using the $S$-matrix  for the imaginary coupling bulk affine
Toda field theories proposed by Hollowood,
and the
other uses a functional integral approach together with a
bootstrap procedure. Evidence to support the results is
collected in various ways: for instance, through the calculation
of the transmission factors for the lightest breathers.
While previous discoveries within  the sine-Gordon model motivated
this study, there are several new phenomena displayed
in the $a_2$ model
including intriguing disparities between the
classical and the quantum pictures. For example, in the
quantum framework, for a specific range of the coupling constant that excludes
a neighbourhood of the classical limit,
there is an unstable bound state.

\vfill
\end{titlepage}

\section{Introduction}

More than a decade ago, Delfino, Mussardo and Simonetti
\cite{Delf94} kindled interest in examining defects in integrable
quantum field theories and since then some progress has been made in
various directions although there remain many open problems. It is
not the purpose of this article to review all the subsequent
developments but a few remarks are in order. The field theories to
be discussed in this paper are non-conformal and describe when
quantised a collection of massive particles. Within a free massive
field theory a defect, for example a defect (or impurity) of $\delta$-function
type, will be accompanied by both transmission and reflection, and
perhaps extra bound states specifically associated with the defect.
However, at least at a classical level, a $\delta$-function defect
within a nonlinear integrable model will destroy integrability. Also
within a quantum field theory containing a defect, the algebraic
constraints to be satisfied by the bulk S-matrix, and the reflection
and transmission factors, as described in \cite{Delf94}\cite{Fring02}
are extremely stringent, and may only be satisfied
with non-zero reflection and transmission provided the bulk S-matrix
is a constant independent of rapidity. Later on,  an alternative
scheme was developed by
Mintchev, Ragoucy and Sorba \cite{Mintchev02}, 
by requiring the
reflection and transmission matrices to satisfy a different algebra. Within this scheme
the S-matrix
need not be trivial in the presence of non-zero reflection and transmission. For
particular quantum field theories - such as the sine-Gordon model, or more generally
any of the affine Toda field theories - with a $\delta$-function
defect,
it remains to be seen
which of these schemes, if indeed either of them,  might turn out to be correct.

\p On the other hand,
one might ask a different question and explore defects that are known to be integrable
within the classical field theory, meaning that they do not destroy classical
integrability, and  subsequently study their role within the corresponding quantum
field theory. This was the approach taken in \cite{bczlandau} and then applied
to a subset of the affine Toda field theories in \cite{bcztoda}. For nonlinear models,
integrable defects, such as those described in \cite{bczlandau}, require discontinuities
in the fields at the location of a defect (rather than discontinuities in their
derivatives, which would be typical of a $\delta$-function discontinuity
 in a nonlinear wave-equation), with specified defect conditions relating the
fields on either side of the defect. For
this reason, they are sometimes called `jump'-defects to emphasise the fact  the
fields are themselves discontinuous. Interestingly, the defect conditions turn out to be
reminiscent of  B\"acklund transformations `frozen' at the site of the defect.
For a recent treatment of these defects and extensions to other models
see \cite{cz06}\cite{caudrelier07}.
Typically, these defects are purely transmitting from a classical point of view
and, for example in the sine-Gordon model, solitons will pass through the defect -
though not unscathed; generally they
will be delayed and might, depending on the precise circumstances, be converted
to an antisoliton, or  be absorbed. Integrable defects studied so far also explicitly break
some or all of the discrete symmetries usually enjoyed by the bulk theories, the
main examples being
parity and time-reversal. This fact implies that solitons travelling from $x<0$ towards
$x>0$ (`left to right') will be affected by the defect in a different manner to those
travelling in the opposite direction.

\p
In a recent article \cite{bczsg05}, devoted to
integrable, purely transmitting defects within the sine-Gordon model, it was
shown how the classical
 defects, introduced in \cite{bczlandau}, may be incorporated
within the associated quantum field theory. In particular, it was demonstrated
how the transmission matrix
discovered originally by Konik and LeClair \cite{Konik97} naturally describes the
behaviour of solitons passing through a defect, with the quantum versions of the
soliton-defect scattering properties matching very closely the classical features.
More precisely, there are
two transmission matrices, one of them labelled by even integers and the other labelled
by odd integers. Alternatively, these may  be described equivalently by the roots and weights of
$a_1$: one of the transmission matrices (with even labels) being labelled naturally
by roots (or integer
spin weights), the other being labelled by the weights of the other
representations (those  of half odd integer spin). It is natural to regard the
transmission matrix labeled by roots
as being unitary (since the sine-Gordon model is a unitary quantum field theory), but then
the transmission matrix labelled by the other weights turns out not to be unitary. In fact,
the states corresponding to the defects described by the latter are unstable soliton-defect
bound states. The appearance in this context of unstable states is an interesting new feature of the
sine-Gordon model. It was also shown how it is possible consistently to allow the classical defects
to move and scatter among themselves. Yet, it remains to be seen how this feature will be realised in
the quantum field theory. Finally, although convincing non-perturbative arguments were provided
for the soliton transmission matrices described in \cite{bczsg05}, it was also shown that breather
transmission matrices are particularly simple and are, at least in principle, amenable to perturbative
calculations.

\p
It is natural to ask if any of these features of integrable defects will emerge in the
imaginary coupling quantum
affine Toda field theories based on  data associated with other algebras. The sine-Gordon
model is the only unitary model within this class of quantum field theories and yet it was pointed out
by Hollowood \cite{Hollowood92} that the classical complex solitons
found within a general affine Toda
field theory have real energy and momentum, and moreover their scattering  might be described by
non-unitary  S-matrices satisfying bootstrap and crossing relations \cite{Hollowood93}.
An assumption made by Hollowood concerned the spectrum of quantum solitons: these are supposed to
be multiplets corresponding to the fundamental representations of the Lie algebra whose
data is used to define a particular affine Toda field theory (for early references, see
\cite{Arinshtein79}). However, a curious feature of the
associated classical field theory is that, apart from the models based on $a_1$ and $a_2$,
the spectrum of classical static solitons is actually different and, in almost all cases, most of the
solutions that should have topological charges corresponding to weights within a fundamental
representation are actually missing; as has been noted by McGhee \cite{McGhee94}. Alternative methods
of constructing solutions \cite{Olive} have not so far revealed the absentees. Presumably, the extra
states in these quantum models are dynamically generated although no detailed mechanism
has been proposed to achieve this. It is tempting to speculate that defects may have something
to do with the story and this idea has provided a partial motivation for this paper.

\section{Jump-defects in the classical $a_r$ affine Toda field theories}

This article will focus on a subset of affine Toda field theories, namely those associated with the
root data of the Lie algebras $a_r$, and in particular of $a_2$. Apart from having the most symmetrical root/weight systems,
these are the models for which classically integrable defects have been described in detail, whose
complex solitons are easy to describe, and whose full set of S-matrices are relatively easy
to calculate using the bootstrap.

\label{classicalATFT}

\p In the bulk, $-\infty <x<\infty$, an affine Toda field theory
corresponding to the root data of the Lie algebra $a_r$ is
 described conveniently by the Lagrangian density
\begin{equation}\label{affineTodaL}
{\cal L} =\frac{1}{2} \, \partial_{\mu}\phi\cdot
\partial^{\mu}\phi-\frac{m^{2}}
 {{\beta}^{2}}\sum_{j=0}^{r}\, (e^{{\beta}\alpha_{j}\cdot\phi}-1),
\end{equation}
where $m$ and ${\beta}$ are constants, and $r$ is the rank of the
algebra. The vectors $\alpha_j$ with $j=1,\dots,r$ are
simple roots (with the convention
$|\alpha_j|^2=2$), and $\alpha_0$ is the lowest root, defined by
$$\alpha_0=-\sum_{j=1}^{r}\,\alpha_j.$$ The field
$\phi=(\phi_1,\phi_2,\, \dots,\, \phi_r)$ takes values in the
$r$-dimensional Euclidean space spanned by the simple roots
$\{\alpha_j\}$. The extra root $\alpha_0$ distinguishes between  the
massive affine and the massless non-affine Toda field theories. The
massive affine theories are  integrable, possessing infinitely many
conserved charges, a Lax pair representation, and many other
interesting properties, both classically and in the quantum domain.
The simplest choice ($r=1$)  coincides with the
sinh-Gordon model. For further details concerning the affine Toda
field theories, see \cite{Arinshtein79}\cite{cdrs} and the review
\cite{Corrigan94}, where further references can be found.

\p After quantisation, provided the coupling constant ${\beta}$ is
real, and the fields are restricted to be real, the $a_r$ affine Toda field theory
describes $r$ interacting scalars, also known as fundamental Toda
particles, whose classical mass parameters are given by
\begin{equation}\label{Todaparticlemasses}
m_a=2\,m\sin\left(\frac{\pi a }h\right), \quad a=1,2\dots,r,
\end{equation}
where $h=r+1$ is the Coxeter number of the algebra. On the other
hand, if the fields are permitted to be complex each affine Toda field theory possesses
classical `soliton' solutions \cite{Hollowood92}. Conventionally,
complex affine Toda field theory are described by the Lagrangian density
\eqref{affineTodaL} in which the coupling constant $\beta$ is
replaced with $i \beta$. Once complex fields are allowed it is clear
that the potential appearing in the Lagrangian density
\eqref{affineTodaL} vanishes whenever the field $\phi$ is constant
and equal to
\begin{equation}\label{}
\phi=\frac{2\pi \,w}{\beta}\quad\mbox{with}\quad \alpha_j\cdot
w\in {\bf Z}, \quad\mbox{i.e.}\quad w\in\Lambda_W(a_{r}),
\end{equation}
where $\Lambda_W(a_{r})$ is the weight lattice of the Lie algebra
$a_r$. These constant field configurations have zero energy and
correspond to stationary points of the affine Toda potential.
Soliton solutions smoothly interpolate between these vacuum
configurations as $x$ runs from $-\infty$ to $\infty$. It is natural
to define the
`topological charges' characterizing such solutions
as follows:
\begin{equation}\label{topologicalcharges}
Q=\frac{\beta}{2\pi}\int^{\infty}_{-\infty}dx\,\partial_x\phi=
\frac{\beta}{2\pi}\left[\phi(\infty,t) -\phi(-\infty,t)\right],
\end{equation}
and these lie in the weight lattice $\Lambda_W(a_r)$. Assuming
$\phi(-\infty,t)=0$, static solitons may be found for which
$\phi(\infty,t)$ lies in a subset of the weight lattice. In
particular, there are static solutions corresponding to weights
within each of the representations with highest weight $w_a,\ a=1,\dots,r$,
satisfying
\begin{equation}
\alpha_i\cdot w_a=\delta_{i a},\quad i,a = 1,\dots, r.
\end{equation}
Explicitly boosted solutions of this type that correspond to the
representation labelled by $a$ have the form
\begin{equation}\label{todasoliton}
\phi^{(a)}=\frac{m^2i}{\beta}\sum^{r}_{j=0}\alpha_{j}\ln \left(1+
E_a\,\omega^{aj}\right), \quad E_a=e^{a_a x-b_a t +\xi_a},\quad
\omega=e^{2\pi i/h},
\end{equation}
where $(a_a, b_a)=m_{a}\, (\cosh{\theta},\sinh{\theta})$, $\xi_a$ is
a complex parameter, and $\theta$ is the soliton rapidity.
Despite the solutions \eqref{todasoliton} being complex, Hollowood \cite{Hollowood92}
showed their total energy and momentum is actually real and requires masses
for  static single solitons
proportional to the mass parameters of the real scalar theory. These are given by
\begin{equation}\label{singlesolitonmasses}
M_a=\frac{2\,h\,m_a}{\beta^2}, \quad a=1,2\dots,r.
\end{equation}
Moreover,
for each $a=1,\dots, r$ there are several solitons
whose topological
charges lie in the set of weights of the fundamental $a^{th}$
representation of $a_r$ \cite{McGhee94}. However, apart from the two extreme
cases, $a=1$ and $a=r$,
not every weight belonging to one of the other representations corresponds to a static
soliton.
The number of possible charges for the representation with label $a$ is exactly equal
to the greatest common divisor of $a$ and $h$, the relevant weights being orbits
of the Coxeter element, and explicit expressions for them may be found in \cite{McGhee94}.
The parameter $\xi_a$ is almost arbitrary but clearly has to be chosen so that there
are no singularities in the solution as $x,t$ vary; shifting $\xi_a$ by $2\pi ia/h$
changes the topological charge. For the two extreme representations (with $a=1$ or
$a=r$), it is clear repeated use of this translation processes the charges through
the full set of weights.

\p The affine Toda field theories  \eqref{affineTodaL} based on $a_r$ generalize the
sinh-Gordon model and the primary purpose of this article is to
extend the techniques and results of recent work devoted to the
sine-Gordon model \cite{bczsg05} to investigate the manner
in which an integrable discontinuity, or `jump' defect, can be accommodated within
the quantum field theory associated with a more general class of field theories. From a purely
classical perspective, the defects have been described before
\cite{bcztoda}. However, for completeness the main features will be
reviewed here together with some additional observations.

\p There are several types of integrable defect for $a_r$ affine Toda field theory
and the distinctions between
them are explained in \cite{bcztoda}. To maintain clarity, most of
the calculations will relate to a specific choice of defect with comments on the other
possiblities relegated to the last section. Bearing this in mind, a single defect located
at $x=0$ may be described by the following modified Lagrangian density 
\begin{equation}\label{affineTodadefectL}
{\cal L}_d=\theta(-x){\cal L}_\phi +\theta(x){\cal
L}_\psi+\delta(x)\left(\half(\phi\cdot E\partial_t\phi +\phi\cdot
D\partial_t\psi - \partial_t\phi\cdot D\psi
    +\psi\cdot E\partial_t\psi) -{\cal B}(\phi , \psi )\right),
\end{equation}
where $E$ is an antisymmetric matrix, $D=1-E$,
\begin{equation}
{\cal L}_\phi =\frac{1}{2} \,\partial_{\mu}\phi\cdot
\partial^{\mu}\phi+\frac{m^{2}}
 {\beta^{2}}\,\sum_{j=0}^{r}\, (e^{i\beta\alpha_{j}\cdot\phi}-1),
\end{equation}
and
\begin{equation}\label{affineTodadefectpotential}
   {\cal B}=-\frac{m}{\beta^2}\sum_{j=0}^r\left(\sigma\,
  e^{i\beta\alpha_j\cdot(D^T\phi+D\psi)/2}+
 \frac{1}{\sigma}\, e^{i\beta\alpha_j \cdot D(\phi
 -\psi)/2}\right).
\end{equation}
Here, $\phi$ and $\psi$ are the fields on the left and on the right
of the defect, respectively, and $\sigma$ is the defect parameter.
The matrix $D$ satisfies the following constraints
\begin{equation}\label{constraintsonD}
\alpha_k\cdot D\alpha_j=\left\{%
\begin{array}{ll}
    \phantom{-}2 & \hbox{$k=j$,} \\
    -2 & \hbox{$k=\pi(j)$,} \\
    \phantom{-}0 & \hbox{otherwise,} \\
\end{array}%
\right.\qquad D+D^T=2,
\end{equation}
where $\pi(j)$ indicates a permutation of the simple roots. Choosing
the `clockwise' cyclic permutation,
\begin{equation}\label{permutation}
\nn \alpha_{\pi(j)}=\alpha_{j-1},\ j=1,\dots, r,\quad
\alpha_{\pi(0)}=\alpha_r,
\end{equation}
the set of constraints \eqref{constraintsonD} is satisfied by the
 choice,
\begin{equation}\label{D}
D=2\sum_{a=1}^r w_a\left( w_a-w_{a+1}\right)^T,
\end{equation}
where the vectors $w_a,\ a=1,\dots ,r$ are the fundamental highest
weights of the Lie algebra $a_r$,
 with the added convention
$w_0\equiv w_{r+1}=0$.  Note, the `anticlockwise' cyclic permutation
used in \cite{bcztoda} is effected by substituting the matrix
\eqref{D} by its transpose.

\p Given the modified Lagrangian density \eqref{affineTodadefectL} the
corresponding equations of motion and defect conditions
are, respectively,
\begin{eqnarray}\label{equationsofmotion}
\partial^{2}\phi=\frac{m^2i}{\beta}\,
\sum_{j=0}^{r}\,\alpha_j\,e^{i\beta\alpha_{j}\cdot\phi}\quad &x<0, \nn \\
\partial^{2}\psi=\frac{m^2i}{\beta}\,
\sum_{j=0}^{r}\,\alpha_j\,e^{i\beta\alpha_{j}\cdot\psi}\quad &x>0,
\end{eqnarray}

\begin{eqnarray}\label{boundaryconditions}
\partial_{x}\phi-E \pt \phi-D \pt \psi+\partial_\phi \cb=0
 \quad &x=0, \nn \\
\partial_{x}\psi-D^{T}\pt \phi +E\pt \psi
-\partial_\psi \cb=0 \quad&x=0.
\end{eqnarray}

\p There are several basic properties of \eqref{boundaryconditions}
that are worth noting. Shifting the fields $\phi,\ \psi$ by roots
yields another solution with the same energy and momentum. This is
because both the bulk and defect potentials are invariant under the
translations
\begin{equation}
\phi\rightarrow \phi +2\pi r/\beta ,\quad \psi\rightarrow
\psi+2\pi s/\beta,
\end{equation}
where $r,s$ are any two elements of the root lattice. In particular,
constant fields
\begin{equation}\label{staticconfigurations}
(\phi,\psi)=2\pi (r,s)/\beta
\end{equation}
 all have the same
energy and momentum despite having a discontinuity at the location
of the defect. Writing $\sigma=e^{-\eta}$, the energy-momentum of
each of these configurations is
\begin{equation}\label{rootroot}
({\cal E}_0,\, {\cal P}_0)=-\frac{2hm}{\beta^2}(\cosh\eta,\
-\sinh\eta ).
\end{equation}
Other constant configurations are possible and, because of the
invariance under translations by roots, it is enough to consider
configurations $(\phi,\psi)=2\pi (w_p,w_q)/\beta$, where $w_p,w_q$
are fundamental highest weights. These are the other
possible constant solutions to \eqref{boundaryconditions}, with
energy-momentum given by
\begin{equation}\label{rootweight}
({\cal E}_a,\, {\cal
P}_a)=-\frac{2hm}{\beta^2}\left[\cosh\left(\eta+\frac{2 a\pi
i}{h}\right),\ -\sinh\left(\eta+\frac{2 a\pi i}{h}\right)\right],\
a=(p-q)\quad p,q=1,\dots, r.
\end{equation}
It is perhaps surprising there is a conserved momentum associated with the defect.
However, that this should be so was pointed out in \cite{bcztoda}, and the expressions
given there
have been used to calculate the above. The expressions in \eqref{rootweight} are
complex, and that is not in itself a surprise, yet all lie on the same mass shell
as \eqref{rootroot}, which is perhaps more surprising.

\p The essential step in calculating \eqref{rootweight} relies on the fact that the
fundamental weights satisfy:
\begin{equation}
w_j\cdot w_p = C^{-1}_{jp},\ \mbox{where}\ \alpha_j \cdot \alpha_p =
C_{jp},\nn
\end{equation}
the latter being the Cartan matrix for $a_r$ (see \cite{Wybourne} for some details concerning
roots and weights). Note, by using \eqref{constraintsonD},
$$\frac{1}{2}\alpha_j \cdot Dw_p= (w_j\cdot w_p)-
(w_{j+1}\cdot w_p),\ \ j=0,\dots,r,
$$
and the explicit form of
the inverse Cartan matrix,
$$C^{-1}=\frac{1}{h}\,\left(\begin{array}{ccccc}r&r-1&r-2& \dots &1\\
                            r-1&2(r-1)&2(r-2)&\dots &2\\
                            r-2&2(r-2)&3(r-2)&\dots&3\\
                            ..&..&..&\dots&..\\
                            ..&..&..&\dots&..\\
                            1&2&3&\dots&r \\\end{array}\right),$$
a direct calculation reveals
$$\frac{1}{2}\alpha_j \cdot Dw_p = \frac{a}{h}\quad j\geqslant p,
\qquad \frac{1}{2}\alpha_j \cdot Dw_p =-\frac{(h-p)}{h}\quad j<p,$$
independently of the label $j$. Similarly,  $(\alpha_j\cdot
D^Tw_p)/2$ can be calculated.

\p The system described by the Lagrangian density
\eqref{affineTodadefectL} is neither invariant under parity nor under
time reversal.  By convention, a soliton with positive rapidity will travel from
the left to the right and, at some time, it will meet the defect
located at $x=0$. The soliton $\psi$ emerging on the right will be
similar to $\phi$, but  delayed. It is  described by,
\begin{equation}\label{solitonsolution}
\psi^{(a)}=\frac{m^2i}{\beta}\sum^{r}_{j=0}\alpha_{j}\ln \left(1+
z_a\,E_a\,\omega^{aj}\right).
\end{equation}
The expression for the delay $z_a$ was derived in \cite{bcztoda} for
the `anticlockwise' permutation. To obtain the delay for the present
situation it is enough to send the $a^{\rm th}$ soliton to the
$(h-a)^{\rm th}$ soliton in the formula appearing in \cite{bcztoda}.
Therefore the delay is given by
\begin{equation}\label{delay}
z_a=\left(\frac{e^{\,-(\theta-\eta)}+i\,e^{-i\gamma_a}}
{e^{\,-(\theta-\eta)}+i\,e^{\,i\gamma_a}}\right), \quad
\gamma_a=\frac{\pi \,a}{h}.
\end{equation}
The delay is generally  complex with  exceptions being
self-conjugate solitons, corresponding to $a=h/2$ (with $r$ odd),
for which the delay is real. In such cases, the delay is equal to the
delay found for the sine-Gordon model \cite{bczlandau}:
\begin{equation}
z=\left(\frac{1+e^{\,-(\theta-\eta)}}{1-e^{\,
-(\theta-\eta)}}\right)= \coth\left(\frac{\theta-\eta}{2}\right).
\end{equation}
Note also that the delays experienced by a soliton, labelled $a$,
and its associated antisoliton, labelled $\bar a = h-a$, are
complex conjugates since $z_{\bar{a}}=\bar{z}_{a}$. For this
reason, solitons and antisolitons are expected to behave
differently as they pass  a defect.

\p The argument of the phase of the delay \eqref{delay} is given by
\begin{equation}\label{delayarg}
\tan(\mbox{arg}\, z_a)=-\left(\frac{\sin
2\gamma_a}{e^{-2(\theta-\eta)} + \cos2\gamma_a}\right),
\end{equation}
implying that the phase shift produced by the defect can vary
between zero (as $\theta\rightarrow -\infty$) and $-2\gamma_a$ (as
$\theta \rightarrow \infty$), decreasing if necessary through
$-\pi/2$ if $\cos2\gamma_a <0$. On the other hand, the boundaries
between the different topological charge sectors in terms of the
imaginary part of $\xi_a$ (eq\eqref{todasoliton}) are separated by
exactly $2\gamma_a$. This means that a soliton might convert  to one
of the adjacent solitons  as it passes the defect provided
$\mbox{arg}\, z_a$ is sufficiently large. In effect, the defect
imposes a rather severe selection rule on the possible topological
charges of the emerging soliton. In the quantised theory, it is
expected that either the transition matrix has zeroes to reflect
this selection rule, or severely suppressed matrix elements to
represent tunnelling between classically disconnected
configurations. In the sine-Gordon model such an effect would not be
noticed  because the basic representation includes just two states
and transitions between them are always permitted.

\p The delay \eqref{delay} diverges when
\begin{equation}\label{boundstate}
\theta = \eta + \frac{i\pi}{2}\left(1-\frac{2a}{h}\right),
\end{equation}
and, with the exception of self-conjugate solitons having $a=h/2$ (including the sine-Gordon
model where $(a,h)=(1,2)$),
this implies a soliton with real rapidity cannot be absorbed by
a defect. For the sine-Gordon model it was noted already that
a classical defect can absorb a soliton and, within the quantum theory,
this phenomenon implies the existence of unstable bound states.
Once the affine Toda field theories are quantised, however, poles
in locations given by \eqref{boundstate} may correspond to additional states
that possess no classical counterpart. The positions of the poles are expected to
depend on the coupling and it might be the case that there is a range of couplings
for which a bound state exists without the range including the classical limit.
It is this fact that suggests that
defects may be part of the explanation for the missing solitons in the
classical models. It will be demonstrated later that a phenomenon rather like this does
actually occur in the $a_2$ model.

\p More generally, the delay \eqref{delay} satisfies a classical bootstrap in the sense that
when two particles $a,b$ in the real quantum field theory have a bound state $\bar c$
the corresponding pole in their S-matrix will occur at rapidities
\begin{equation}
\theta_a=\theta_c-i\bar{U}_{ac}^b,\quad \theta_b= \theta_c +i\bar{U}_{bc}^a,
\end{equation}
and the corresponding delays \eqref{delay} in the complex classical theory satisfy
\begin{equation}
z_a(\theta-i\bar{U}_{ac}^b)\, z_b(\theta +i\bar{U}_{bc}^a)=z_{\bar c}(\theta).
\end{equation}
This is not difficult to check directly using the $a_r$ coupling data \cite{cdrs}.

\p All these observations, and the experience gained with the sine-Gordon model,
suggest the investigation of the corresponding quantum theory should be
interesting even in the next simplest $a_2$ model.

\section{The fundamental $S$-matrices for the $a_r$ affine Toda field theories}
\label{Smatrix}

\p The $S$-matrices describing the scattering of solitons in the
$a_r$ affine Toda field theory were conjectured by Hollowood \cite{Hollowood93}.
Hollowood's proposal makes use of the $R$-matrices of the quantum
group $U_q(a_r)$, specifically the trigonometric solutions of the
Yang-Baxter equation (YBE) initially found by Jimbo \cite{Jimbo89} (and references therein).
The basic assumption asserts that the particles of the $a_r$ affine Toda field theory
lie in the $r$ different multiplets corresponding to the $r$
fundamental representations of $U_q(a_r)$. The $S$-matrix $S^{ab}$
describing the scattering of two particles with rapidities
$\theta_1$ and $\theta_2$, lying in the multiplets $a$ and $b$,
respectively, is an interwining map on the two representation spaces
$V_a$ and $V_b$. In other words,
\begin{equation}
S^{ab}(\theta_{12}): V_a\otimes V_b \rightarrow V_b \otimes
V_a,\quad \theta_{12}=(\theta_1-\theta_2),
\end{equation}
and $S$ has the following form
\begin{equation}
S^{ab}(\theta_{12})=\rho^{ab}(\theta_{12})\,R^{ab}(\theta_{12}),
\end{equation}
where $R^{ab}$ is a $U_q(a_r)$ $R$-matrix and $\rho^{ab}$  is a
scalar function determined by the requirements of `unitarity',
crossing symmetry, analyticity and consistency relations (bootstrap
constraints), which a scattering matrix must satisfy
\cite{Hollowood93}.

\p For the purposes of the present article, explicit expressions for
the $S$-matrices are needed. In particular, for the $a_r$ affine Toda field theory the
explicit expression of the $S$-matrix describing the scattering of
the solitons in the first representation, namely the matrix $S^{11}$
also referred to as the fundamental scattering matrix, will be
provided in this section.

\p The representation space $V_1$ of the first multiplet has
dimension $h$ and its states are the solitons $A^1_j$,
$j=1,\dots,h$. The weights of this representation are conveniently described
by  \cite{Wybourne}
\begin{equation}\label{weights1}
l_{j}^1\equiv
l_j=\sum_{l=1}^r\,\frac{(h-l)}{h}\alpha_l-\sum_{l=1}^{j-1}\,\alpha_l,
\quad j=1,\dots,h.
\end{equation}

\p The elements of  $S^{11}$ can be described conveniently
using the non-commutative Faddeev-Zamolodchikov algebra. Consider
$A^1_j$ ($j=1,\dots,h$) to be  generators of such an algebra.
Then, the non-zero elements of $S^{11}$ represent the following
relations processes \cite{Hollowood93,Gandenberger95} for
$j,k=1,\dots,h$ and $\theta_{12}>0$
\begin{eqnarray}\label{Smatrixcommrelation}
A^1_j(\theta_1)A^1_j(\theta_2)&=&S^{11}\,{^{jj}_{jj}}\,(\theta_{12})
\, A^1_j(\theta_2)A^1_j(\theta_1),\nn\\
A^1_j(\theta_1)A^1_k(\theta_2)&=&S^{11}{\,^{kj}_{jk}}\,(\theta_{12})
\, A^1_k(\theta_2)A^1_j(\theta_1) +S^{11}{\,
^{jk}_{jk}}\,(\theta_{12})\, A^1_j(\theta_2)A^1_k(\theta_1), \quad
j\neq k,
\end{eqnarray}
with
\begin{eqnarray}\label{S11elements}
S^{11}\,{^{jj}_{jj}}\,(\theta_{12})&=&\rho^{11}(\theta_{12})\,
\left(q\,x_{12}-q^{-1}\,x_{12}^{-1}\right),
\nn\\
S^{11}\,{^{kj}_{jk}}\,(\theta_{12})&=&\rho^{11}(\theta_{12})\,
\left(x_{12}-x_{12}^{-1}\right), \quad k\neq j, \nn\\\nn\\
S^{11}\,{^{jk}_{jk}}\,(\theta_{12})&=&\rho^{11}(\theta_{12})\,\left(q-q^{-1}\right)\left\{%
\begin{array}{ll}
x_{12}^{\phantom{-}(1-2|l|/h)}\left.\right|_{\,l=j-k<0}\\
\\
x_{12}^{-(1-2|l|/h)}\left.\right|_{\,l=j-k>0}\\
\end{array}%
\right.
\end{eqnarray}
and
\begin{equation}
\quad x_j=e^{h\gamma \theta_j/2},\quad j=1,2;\quad
x_{12}=\frac{x_1}{x_2}; \quad q=-e^{-i\pi\gamma}, \quad
\gamma=\frac{4\pi}{\beta^2}-1.
\end{equation}
The function $\rho^{11}$ is given by the following expression
\cite{Hollowood93}:
\begin{eqnarray}\label{rhofunction}
\rho^{11}(\theta_{12})&=&
\,\,\frac{\Gamma(1+h\gamma i\,\theta_{12}/2\pi)
\Gamma(1-h\gamma i\,\theta_{12}/2\pi-\gamma)}{2\pi i}\,\,
\frac{\sinh(\theta_{12}/2+i\pi/h)}{\sinh(\theta_{12}/2-i\pi/h)}\nn\\
&&\qquad\qquad \times\ \prod_{k=1}^\infty\,
\frac{F_k(\theta_{12})\,F_k(2\pi i/h-\theta_{12})}{F_k(2\pi i
/h+\theta_{12})\, F_k(2\pi i -\theta_{12})},
\end{eqnarray}
where
\begin{equation}\label{}
    F_k(\theta_{12})=\frac{
    \Gamma(1+h\gamma i\,\theta_{12}/2\pi+hk\gamma)}
    {\Gamma(h\gamma i\,\theta_{12}/2\pi+(hk+1)\gamma)
    }.
\end{equation}

\p In principle,  $S^{11}$ is enough to describe the
quantum affine Toda field theory since the remaining $S$-matrices for solitons in the
other fundamental representations can be determined by adopting a
bootstrap procedure. Most expressions for the remaining
$S$-matrices are neither needed nor provided, though a description
of the soliton states $A_j^a$ in the representation $a$ in terms of
states $A_j^1$ in the first representation will be given and used in
the next section. The scattering matrix $S^{12}$ for $a_2$ will be used
later and is provided in Appendix A.

\p The bootstrap linking states lying in two
different representations is given schematically
\begin{equation}
A^c_i(\theta)\equiv \sum_{j,k}\,c_i{^{jk}}\,A^a_j(\theta-\theta_p/2)\;
A^b_k(\theta+\theta_p/2),\quad l^c_i = l^a_j+l^b_k,
\end{equation}
where $\theta_p$ is the location of the pole in the scattering
matrix $S^{a b}$ corresponding to a soliton in the representation
labelled $c$. For instance, starting from the operator $A_j^1$, for
which the scattering $S$-matrix is known, the solitons in the second
representation will be represented by
\begin{eqnarray}\label{solitons2}
A^2_i(\theta)\equiv \sum_{j,k}\,c_i{^{jk}}\,A^1_j(\theta-i\pi/h)\;
A^1_k(\theta+i\pi/h),\quad l^2_i= l^1_j+l^1_k,\nn\\
c_i{^{jk}}=(-q)^{-(1-2|j-k|/h)}\, c_i{^{kj}},\quad
j<k=1,\dots,h.
\end{eqnarray}
Note, each weight in the second representation can be expressed in only one
way as a sum of weights in the first representation. Hence, the sum in
\eqref{solitons2} contains just two terms related as shown.
Iterating this process allows a formal presentation for all the states
in each fundamental representation.

\section{Functional integral approach to the transmission matrix}
\label{functionalapproach}

Before considering in detail all solutions to the triangle
equations that express the compatibility between the bulk $S$-matrix
and the transmission matrix - and bearing in mind
there are likely to be several formal solutions to the triangle equations, not
all of which might be relevant to the present problem - it is worth
extending the functional integral argument introduced in
\cite{bczsg05}. This will supply some constraints that will be helpful in
discriminating among the variety of solutions. In particular, the
functional integral allows a comparison between the elements of
the transmission matrix describing the evolution of field
configurations in the presence of a defect labelled by a pair of
roots $(r,s)$ and the evolution of field configurations in the
presence of the defect labelled by $(0,0)$. The basic idea
is to shift the fields by setting
$$\phi\rightarrow \phi-\frac{2\pi r}{\beta}, \quad
\psi\rightarrow \psi-\frac{2\pi s}{\beta},$$ and use the
invariance of the bulk action and the defect potential. The
remaining pieces of \eqref{affineTodadefectL}, the terms linear in
time derivatives, lead to the expression
\begin{equation}
T(r,s)=e^{i\tau(r,s)}\, T(0,0),
\end{equation}
where
\begin{equation}
\tau(r,s)=\frac{\pi}{\beta}\left(-\delta\phi\cdot
(Er+Ds)+(rD+sE)\cdot\delta\psi\right),
\end{equation}
and $\delta\phi,\ \delta\psi$ are the changes in the field
configurations from  initial to  final states.

\p A soliton passing the defect will either retain its topological
charge $\lambda$, or its charge will change to $\mu$, one of the other
weights within the representation to which the soliton belongs. Thus, the effect of a
soliton passing a defect must be to change the defect  labels by
\begin{equation}
r\rightarrow r-\lambda,\quad s\rightarrow s-\mu,
\end{equation}
and, therefore,
$$\delta\phi=-\frac{2\pi \lambda}{\beta}, \quad \delta\psi=
-\frac{2\pi \mu}{\beta}.$$ Thus,
\begin{equation}\label{Deltafirst}
\tau(r,s)= \frac{2\pi^2}{\beta^2}\left(\lambda\cdot
(Er+Ds)-(rD+sE)\cdot \mu\right),
\end{equation}
which is written more conveniently (using $D=1-E$) as
\begin{equation}
\tau(r,s)=
\frac{2\pi^2}{\beta^2}\left(\frac{1}{2}(\lambda-\mu)\cdot(r+s)
-(\lambda-\mu)\cdot
E(s-r)+\frac{1}{2}(\lambda+\mu)\cdot(s-r)\right).
\end{equation}
In other words, using this argument it is expected that
\begin{equation}
T(r,s,\lambda,\mu)=Q^{[(\lambda-\mu)\cdot p -2 (\lambda-\mu)\cdot
E \alpha+(\lambda+\mu)\cdot \alpha]/4}\, T(0,0,\lambda,\mu),\quad
Q\equiv e^{4\pi^2 i/\beta^2}=q^{-1},
\end{equation}
where $p=s+r$ and $\alpha=s-r$. Naturally, this style of argument
can give no information concerning the rapidity dependence of the
transmission matrix but it does suggest, assuming the conservation
of topological charge, that a general element of the transmission
matrix should have the form:
\begin{equation}\label{Tgeneral}
T_{\lambda \alpha p}^{\mu\beta q}(\theta)= Q^{[(\lambda-\mu)\cdot p
-2 (\lambda-\mu)\cdot E \alpha+(\lambda+\mu)\cdot \alpha]/4}\,
T_{\lambda}^\mu (\theta)\, \delta_{\alpha}^{\beta-\lambda+\mu}\,
\delta_p^{q+\lambda+\mu}
\end{equation}

\p Also, the dependence on $p$ can be eliminated  using the  unitary
transformation
\begin{equation}\label{unitarytran}
U_{\alpha r}^{\beta s}=Q^{\alpha\cdot
r/4}\delta_{\alpha}^{\beta}\delta_r^s,
\end{equation}
to find:
\begin{equation}\label{T}
 T_{\lambda\alpha p}^{\mu\beta q}(\theta)= Q^{[2\alpha\cdot
E(\lambda-\mu)+(\alpha+\lambda)^2-(\alpha-\mu)^2]/4} \,
T_\lambda^\mu(\theta)\, \delta_{\alpha}^{\beta-\lambda+\mu}\,
\delta_p^{q+\lambda+\mu}.
\end{equation}
For the fundamental representations of $a_r$, labelled $a=1$ or $r$,
the weights have equal length and the expression simplifies a
little.
Thus, for solitons in the first representation, equation \eqref{T}
simplifies to
\begin{equation}\label{T1}
T^1{_{i\alpha p}^{j\beta q}}(\theta)= Q^{\,\alpha\cdot
[E(l_i-l_j)+l_i+l_j]/2} \, T^1{_{i}^{j}}(\theta)\,
\delta_{\alpha}^{\beta-l_i+l_j}\, \delta_p^{q+l_i+l_j},
\end{equation}
where the general weights $\mu,\lambda$ have been replaced by the
weights lying in the first representation, namely $l^1_i\equiv l_i$
(see \eqref{weights1}).
Further, when these specific  weights are used as labels, the
notation $T^1{_i^j}$ is used as a simplification. A similar
expression holds for the transmission matrix for solitons in the
representation $r$; for these, the topological charges are merely a
sign change relative to those in the first representation
($l^r_k\equiv l_{\bar{k}} =-l^1_k$). For the case of $a_1$, or the
sine-Gordon model, $E=0$ and \eqref{T1} agrees with the findings of
\cite{bczsg05}.

\p
Before proceeding to solve the triangle equations for the $a_2$
model, it is  instructive  to apply the bootstrap procedure to
 the general form of $T^1$, given in  \eqref{T1},  to see the extent to which it
is possible, solely from the bootstrap, to gather information about
the classical quantity $E$ and the still undetermined part of the
transmission matrix. For this purpose, consider $D_{\alpha}$ to be
the defect operator. Then, it is formally possible to describe the
interaction between a defect and a soliton within the first
fundamental representation as follows ($\theta>0$),
\begin{equation}\label{defectoperator}
A^1_{i}(\theta)D_{\alpha}=T^1{^{j\beta}_{i\alpha}}(\theta)D_{\beta}A^1_{j}(\theta).
\end{equation}
Note, the indices $p$ and $q$ do not appear in
\eqref{defectoperator} since, as already established, the transmission
matrix does not depend on them. Note also that \eqref{defectoperator}
is consistent with the notation \eqref{Smatrixcommrelation} used for
the $S$-matrix. The interaction
between the defect and  solitons in the second representation will be represented by
\begin{eqnarray}\label{boostrap1to2}
A^2_i(\theta)\,D_\alpha
&=& T^2{^{m\beta}_{i\alpha}}(\theta)D_{\beta}A^2_m(\theta)\nn\\
&=&c_i{^{jk}}\,T^1{^{a\delta}_{j\beta}}(\theta -i\pi/h)\,
T^1{^{b\beta}_{k\alpha}}(\theta+i\pi/h)\,
\,D_\delta\,A^1_a(\theta-i\pi/h)\,
A^1_b(\theta+i\pi/h)\nn\\
&=& T^2{^{n\delta}_{i\alpha}}(\theta)\,c_n{^{ab}}\,D_{\delta}A^1_a(\theta-i\pi/h)\,
A^1_b(\theta+i\pi/h)
\end{eqnarray}
where all repeated indices are summed. Thus,
\begin{equation}\label{Stwomatrix}
T^2{^{n\delta}_{i\alpha}}(\theta)\,c_n{^{ab}}=c_i{^{jk}}\,
T^1{^{a\delta}_{j\beta}}(\theta -i\pi/h)\,
T^1{^{b\beta}_{k\alpha}}(\theta+i\pi/h).
\end{equation}

\p Bearing in mind the result obtained for  the transmission matrix
in the sine-Gordon model \cite{bczsg05}, and also noting the
rapidity dependence within the S-matrix \eqref{S11elements}, a suitable
ansatz to adopt for the rapidity independent part of \eqref{T1} is
\begin{equation}
T^1{_{i}^{j}}(\theta)=t_{ij}\,x^{\epsilon_{ij}}\,g^1(\theta),
\end{equation}
where $t_{ij}$ and  $\epsilon_{ij}$ are constants, and
$g(\theta)$ is independent of the soliton labels.

\p
When $a=b$,  the right hand side of \eqref{Stwomatrix} must vanish since there are no
weights of the form $2l^1_a$ in the second representation. As a
consequence, the following relations must hold
\begin{equation}\label{bootstrapconstraint1}
Q^{\,l_k\cdot El_j+l_a\cdot(1+E)(l_k-l_j)}\,c_i{^{jk}}=-(-)^{\epsilon_{ka}-\epsilon_{ja}}
Q^{\epsilon_{ja}-\epsilon_{ka}} \,c_i{^{kj}}.
\end{equation}
Putting $j$  equal to the index $a$, and using the fact (deducible from \eqref{weights1})
that $l_a\cdot(l_a-l_k)=1$ for all $k\ne a$, \eqref{bootstrapconstraint1} becomes
\begin{equation}
c_i{^{ak}}=-(-)^{\epsilon_{ka}-\epsilon_{aa}}Q^{1+\epsilon_{aa}-\epsilon_{ka}}\, c_i{^{ka}},
\end{equation}
and this can be compared with \eqref{solitons2}. Firstly, interchanging $a$ and $k$
requires
\begin{equation}\label{autoconstraints}
\epsilon_{ak}+\epsilon_{ka}-\epsilon_{kk}-\epsilon_{aa}=2, \quad a\ne k.
\end{equation}
Secondly, using \eqref{solitons2} leads to more detailed information, namely,

\begin{equation}\label{exponentsconstraints1}
\epsilon_{ka}=\epsilon_{aa}+2|a-k|/h,\quad a<k\,.
\end{equation}
Together, \eqref{autoconstraints} and \eqref{exponentsconstraints1} determine
 all the off-diagonal exponents appearing in the
transmission matrix in terms of its diagonal exponents.
Also, when $j<k\neq a$ it is possible to gather information
concerning the matrix $E$ because \eqref{bootstrapconstraint1}
demands
\begin{equation}\label{Ematrixconstraints}
(l_k-l_a)\cdot E(l_j-l_a)=-1,\ j<k<a \,\,\mbox{or}\,\, a<j<k\,;\quad
(l_k-l_a)\cdot E(l_j-l_a)=1,\ j<a<k.
\end{equation}
Because of equivalences it is sufficient to consider only one of these sets of relations.
Making use of \eqref{weights1}, the
constraints implied by \eqref{Ematrixconstraints} can be rewritten
\begin{equation}\label{Ematrixconstraints+}
(\alpha_k+\alpha_{k+1}+\cdots+\alpha_{a-1})\cdot
E(\alpha_j+\alpha_{j+1}+\cdots+\alpha_{k-1})=-1,\ j<k<a.
\end{equation}
The independent relations provided by \eqref{Ematrixconstraints+}
state the following
\begin{equation}\label{Ematrix}
\alpha_l\cdot E\alpha_m=\left\{%
\begin{array}{ll}
    -1 & \hbox{$m=l-1$,} \\
    \phantom{-}0 & \hbox{$m=1,\cdots l-2$,} \\
\end{array}%
\right.\qquad l=2,\cdots,a-1.
\end{equation}
Since the index $a$ takes the values= $1,\dots,h$, the total number of
independent constraints in
\eqref{Ematrix} is $r(r-1)/2$ and precisely equal to the number of
degrees of freedom of the matrix $E$. Consequently, $E$ is
completely determined by the bootstrap procedure and can be compared
with the formula \eqref{constraintsonD} which defined the matrix
$E=1-D$ established in the classical setting of the defect problem. It can
be seen that the two expressions coincide, provided the clockwise
cyclic permutation of the simple roots in formula
\eqref{constraintsonD} is chosen.

\p Next, consider the terms for which  $a\neq b$ and $l^2_i=l^1_a+l_b^1$
in \eqref{Stwomatrix}.
Then, $n=i$ and the left hand side of \eqref{Stwomatrix} can be written
in two ways according
to the choice of ordering  $a$ with respect to $b$. Thus,
\begin{eqnarray}
T^2{^{i\gamma}_{i\alpha}}&=&T^1{^{a\gamma}_{a\beta}}(\theta-i\pi/h)
T^1{^{b\beta}_{b\alpha}}(\theta+i\pi/h)
+({c_i^{ba}}/{c_i^{ab}})\,T^1{^{a\gamma}_{b\beta}}(\theta-i\pi/h)T^1{^{b\beta}_{a
\alpha}}(\theta+i\pi/h)\nn\\
&=&T^1{^{b\gamma}_{b\beta}}(\theta-i\pi/h)T^1{^{a\beta}_{a\alpha}}(\theta+i\pi/h)
+({c_i^{ab}}/{c_i^{ba}})\,T^1{^{b\gamma}_{a\beta}}(\theta-i\pi/h)T^1{^{a\beta}_{b
\alpha}}(\theta+i\pi/h).
\end{eqnarray}
Since the right hand sides must match, and since the dependence on $\theta$ is different in
different terms, there is an additional pair of constraints. Specifically, these are
\begin{equation}\label{exponentsconstraints2}
\epsilon_{aa}=\epsilon_{bb}\equiv \epsilon; \quad
\epsilon_{ab}-\epsilon_{ba}=2(1-2|a-b|/h), \quad a<b=1,\dots,h.
\end{equation}
Therefore, all the diagonal exponents in the
transmission matrix are the same; the relations in the second
group are precisely the differences of the relations given earlier
in \eqref{exponentsconstraints1}. Using
\eqref{exponentsconstraints1} and \eqref{exponentsconstraints2} the
diagonal terms of the transmission matrix for the solitons in the
second representation are
\begin{equation}
T^2{^{i\gamma}_{i\alpha}}(\theta)=Q^{\alpha\cdot(l_a+l_b)}
\,x^{2\epsilon}(t_{aa}t_{bb}+x^2
t_{ab}t_{ba})\,
g^1(\theta-i\pi/h)\,g^1(\theta+i\pi/h)\, \delta_{\alpha}^{\gamma},
\quad l^2_i=l_a+l_b.
\end{equation}

\p It is possible to go a little further in the analysis of
\eqref{Stwomatrix} by looking at those cases for which
\begin{equation}\label{constraintijab}
\frac{c_i{^{jk}}}{c_i{^{kj}}}=\frac{c_n{^{ab}}}{c_n{^{ba}}}.
\end{equation}
Since $l^2_n$ is uniquely $l^1_a+l^1_b$ there is only the choice of
ordering $a$ and $b$ when considering \eqref{Stwomatrix}. Because of
this, and the fact $c_n{^{ab}}\ne c_n{^{ba}}$, there must be further
constraints on $T^1(\theta)$. With the particular choice
\eqref{constraintijab}, these are
\begin{eqnarray}\label{constraintijabapplied}
&&T^1{^{a\delta}_{j\beta}}(\theta -i\pi/h)\,
T^1{^{b\beta}_{k\alpha}}(\theta+i\pi/h)+\frac{c_i{^{kj}}}{c_i{^{jk}}\,}
\,T^1{^{a\delta}_{k\beta}}(\theta
-i\pi/h)\, T^1{^{b\beta}_{j\alpha}}(\theta+i\pi/h)\nn\\
&&\ \ \  =\frac{c_i{^{jk}}}{c_i{^{kj}}\,}\,T^1{^{b\delta}_{j\beta}}(\theta
-i\pi/h)\,
T^1{^{a\beta}_{k\alpha}}(\theta+i\pi/h)+T^1{^{b\delta}_{k\beta}}(\theta
-i\pi/h)\, T^1{^{a\beta}_{j\alpha}}(\theta+i\pi/h).
\end{eqnarray}
For definiteness, suppose that $j<k$. Then, the constraint
\eqref{constraintijab} is satisfied (using \eqref{solitons2})
provided $|a-b|=|j-k|$ (if $a<b$), or $|a-b|=h-|j-k|$ (if
$a>b$). The full set of possibilities will not be analysed here and
to illustrate some important points only the
 two simplest cases, namely $|j-k|=1$ and
$|j-k|=h-1$, will be considered in detail. Besides, these cover all
the possibilities for $a_2$. Bearing in mind
that
\begin{equation}
l^2_i=l_j+l_k,\quad l^2_n=l_a+l_b,
\end{equation}
it is useful to list explicitly the combinations of indices
$j,k,a,b$ which will be investigated. Firstly, consider $|j-k|=1$.
For the $a_2$ case such combinations are
\begin{eqnarray}\label{indexcombinationsa2}
&l^2_i=l_1+l_2,\quad l^2_n=l_2+l_3\quad \mbox{if}\quad a<b;\quad
&l^2_i=l_2+l_3,\quad
l^2_n=l_3+l_1\quad\mbox{if}\quad a>b\nn\\
&l^2_i=l_2+l_3,\quad l^2_n=l_1+l_2\quad \mbox{if}\quad a<b;\quad
&l^2_i=l_1+l_2,\quad l^2_n=l_3+l_1\quad \mbox{if}\quad a>b,
\end{eqnarray}
while their generalizations for the $a_r$ affine Toda field theory are
\begin{eqnarray}
&l^2_i=l_j+l_{j+1},\quad l^2_n=l_{j+1}+l_{j+2}\quad \mbox{if}\quad
a<b;\quad &l^2_i=l_{h-1}+l_h,\quad
l^2_n=l_h+l_1\quad\mbox{if}\quad a>b\nn\\
\label{indexcombinationsar1}\\
&l^2_i=l_j+l_{j+1},\quad l^2_n=l_{j-1}+l_j\quad \mbox{if}\quad
a<b;\quad &l^2_i=l_1+l_2,\quad l^2_n=l_h+l_1\quad \mbox{if}\quad
a>b.\nn\\
\label{indexcombinationsar2}
\end{eqnarray}
It should be emphasised that while \eqref{indexcombinationsa2}
represents all possible combinations of indices for $a_2$ with the
constraint $|j-k|=1$, \eqref{indexcombinationsar1} \eqref{indexcombinationsar2}
only provides a subset
of the possibilities for $a_r$. Using \eqref{autoconstraints},
\eqref{exponentsconstraints1}, \eqref{Ematrixconstraints},
\eqref{exponentsconstraints2}, it can be verified that equation
\eqref{constraintijabapplied} is an identity for the weights
in  \eqref{indexcombinationsar1}.  The corresponding transmission matrix
elements for solitons in the second representation are
\begin{eqnarray}
T^2{^{n\delta}_{i\alpha}}(\theta)&=&Q^{\alpha\cdot[E(l_j-l_{j+2})+2l_{j+1}+l_j+l_{j+2}]/2}
\,x^{2(\epsilon+1-2/h)}(t_{jj+2}t_{{j+1
j+1}}+x^2 t_{j+1j+2}t_{jj+1})\nn\\
&&\ \ \ \ \times\, g^1(\theta-i\pi/h)\,g^1(\theta+i\pi/h)\,
\delta_{\alpha}^{\delta-l_j+l_{j+2}},
\end{eqnarray}
with
$$l^2_i=l_j+l_{j+1},\qquad l^2_n=l_{j+1}+l_{j+2}\qquad
j=1,\dots h-2$$ and
\begin{eqnarray}
T^2{^{n\delta}_{i\alpha}}(\theta)&=&Q^{\alpha\cdot[E(l_{h-1}-l_1)+2l_{h}+l_{h-1}+l_1]/2}
\,x^{2(\epsilon+1-2/h)}(t_{h-11}t_{{hh}}+x^2 t_{h1}t_{h-1h})\nn\\
&&\ \ \ \ \times\, g^1(\theta-i\pi/h)\,g^1(\theta+i\pi/h)\,
\delta_{\alpha}^{\delta-l_{h-1}+l_1},
\end{eqnarray}
with
$$ l^2_i=l_{h-1}+l_h,\qquad l^2_n=l_h+l_1.$$ Alternatively, using the
index combinations in
\eqref{indexcombinationsar2}, the expression
\eqref{constraintijabapplied} is satisfied provided the following
constraints on the constants $t_{ij}$ hold
\begin{equation}\label{constanttconstraints1}
t_{jj} t_{j+1j-1}=t_{j+1j} t_{jj-1}\qquad j=2,\cdots,h-1;\qquad
t_{11} t_{2h}=t_{21} t_{1h},
\end{equation}
and the corresponding elements of the transmission matrix for the
solitons in the second representations are equal to zero.

\p Finally,  the case $|i-j|=h-1$ corresponds to the following two-index
combinations for $a_2$ (the possibility $a<b$ having  been
investigated already):
\begin{eqnarray}\label{indexcombinationsa2h-1}
&l^2_i=l_1+l_3,\quad l^2_n=l_2+l_1\quad \mbox{if}\quad a>b,\nn\\
&l^2_i=l_1+l_3,\quad l^2_n=l_3+l_2\quad \mbox{if}\quad a>b;
\end{eqnarray}
and these  generalize for the $a_r$ affine Toda field theory to
\begin{eqnarray}\label{indexcombinationsarh-11}
&l^2_i=l_1+l_h,\quad l^2_n=l_2+l_1\quad \mbox{if}\quad a>b,\\
&l^2_i=l_1+l_h,\quad l^2_n=l_h+l_{h-1}\quad \mbox{if}\quad a>b.\label{indexcombinationsarh-12}
\end{eqnarray}
Just as in  the previous case, using the weights in
 \eqref{indexcombinationsarh-11}, the expression
\eqref{constraintijabapplied} is an identity with the corresponding
transmission matrix element given by
\begin{equation}
T^2(\theta)^{n\,\beta}_{i\,\alpha}=
Q^{\alpha\cdot[E(l_h-l_1)+2l_1+l_2+l_h]/2}
\,x^{2(\epsilon+1-2/h)}(t_{11} t_{h2}+x^2 t_{h1} t_{12})\,
g^1(\theta-i\pi/h)\,g^1(\theta+i\pi/h)\,
\delta_{\alpha}^{\beta-l_h+l_2},
\end{equation}
with $$l^2_i=l_1+l_h,\qquad l^2_n=l_2+l_1.$$ On the other hand, using
the index combination in
\eqref{indexcombinationsarh-12}, the expression
\eqref{constraintijabapplied} forces the following constraint on the
constants $t_{ij}$,
\begin{equation}\label{constanttconstraints2}
t_{hh} t_{1h-1}=t_{1h} t_{hh-1},
\end{equation}
with the corresponding transmission matrix elements being equal to zero.

\p In summary, this partial analysis of the bootstrap procedure
determines the matrix $T^1$ for all the affine Toda field theories up to a
function $g(\theta)$ that is independent of the soliton labels, and up to
constants $t_{ij}$, which are themselves constrained.
Moreover, it has been noted that provided the initial $T^1$ matrix
has all entries different from zero, the transmission matrix $T^2$
is required to have at least some off-diagonal entries equal to zero.
For the simplest case of $a_2$, the analysis based on the
bootstrap  has been carried out completely, and therefore
it is possible to write down the full $T^2$ matrix for
antisolitons.

\p
To conclude, the transmission matrices for the $a_2$ affine Toda field theory
predicted by the bootstrap procedure are

\begin{eqnarray}\label{solitonsTmatrix}
&&T^1\,{^{n\beta}_{i\alpha}}(\theta)=g^1(\theta)\left(
\begin{array}{ccc}
 t_{11}\,Q^{\alpha\cdot l_1}\,\delta_{\alpha}^{\beta}
  & t_{12}\,x^{4/3}\,\delta_{\alpha}^{\beta-\alpha_1}\,
 & t_{13}\,x^{2/3}\,Q^{-\alpha\cdot l_2}
  \, \delta_{\alpha}^{\beta+\alpha_0}\phantom{} \nn\\
  t_{21}\,x^{2/3}\,Q^{-\alpha\cdot l_3}
  \,\delta_{\alpha}^{\beta+\alpha_1}\phantom{}
  &
  t_{22}\,Q^{\alpha\cdot l_2}\,\delta_{\alpha}^{\beta}
  & t_{23}\,x^{4/3}\,\delta_{\alpha}^{\beta-\alpha_2}
   \\
 t_{31}\,x^{4/3}\,\delta_{\alpha}^{\beta-\alpha_0}
& t_{32}\,x^{2/3}\,Q^{-\alpha\cdot l_1}
  \,\delta_{\alpha}^{\beta+\alpha_2}\phantom{}
  & t_{33}\,Q^{\alpha\cdot l_3}
  \,\delta_{\alpha}^{\beta} \\
\end{array}
\right), \nn\\
&&\\&&\nn\\
&&T^2\,{^{\bar{n}\beta}_{\bar{i}\alpha}}(\theta)=g^1(\theta-i\pi/3)\;g^1(\theta+i\pi/3)\;
\left(1+x^2\frac{t_{21} t_{31}t_{13}}{t_{11}
t_{22}t_{33}}\right)\label{antisolitonsTmatrix}\nn\\
&&\nn\\
&&\phantom{aaaaaaaaaaaa}\times\, \left(
\begin{array}{ccc}
 t_{11}\,Q^{-\alpha\cdot l_1}\,\delta_{\alpha}^{\beta}\phantom{}
  & t_{21}t_{33}\,x^{2/3}\,\delta_{\alpha}^{\beta+\alpha_1}
  & 0\\
  0
  &
  t_{22}\,Q^{-\alpha\cdot l_2}\,\delta_{\alpha}^{\beta}\phantom{}
  & t_{32}t_{11}\,x^{2/3}\,\delta_{\alpha}^{\beta+\alpha_2}
 \\
 t_{13}t_{22}\,x^{2/3}\,\delta_{\alpha}^{\beta+\alpha_0}
 \,\phantom{a}
  & 0
  & t_{33}\,Q^{-\alpha\cdot l_3}
  \,\delta_{\alpha}^{\beta}\phantom{} \\
\end{array}
\right),
\end{eqnarray}
with
\begin{equation}\label{tconstants}
\frac{t_{21}\,t_{13}}{t_{11}}=t_{23},\quad
\frac{t_{32}\,t_{21}}{t_{22}}=t_{31},\quad
\frac{t_{32}\,t_{13}}{t_{33}}=t_{12}.
\end{equation}
Note, the function
$g^1(\theta)$ has been redefined in order to absorb the factor
$x^\epsilon$.

\p
At first sight the imbalance between solitons and antisolitons appears
strange and one might wonder about its consistency since the bootstrap
could be run the other way to define $T^1$ starting with $T^2$.
Although the details will not be given here the results are entirely consistent;
starting with a matrix containing these zeroes and using it to define $T^1$ does indeed
recover \eqref{solitonsTmatrix}.

\p
In the next section it will be shown that the transmission matrix
\eqref{solitonsTmatrix} coincides with a solution of the triangle equations.

\section{The transmission matrix for the $a_2$ affine Toda field theory: the
triangle equations} \label{section5}

\p In this section arguments will be restricted to the special case
$a_2$. In order to find
the transmission matrices describing the interaction between the
jump-defect and  solitons, the general procedure
applied successfully for the sine-Gordon model in \cite{bczsg05}
will be adopted. The first step is to solve the triangle
equations, which relate the elements of the transmission matrix to $S$-matrix
elements; and, in the first instance, attention will be
focused on the transmission matrix for solitons in the $a=1$ representation.\footnote{\ From
now on, these solitons will be called simply solitons, while the
solitons in the $a=2$ representation will be called antisolitons.}
Consequently, the triangle equation
reads
\begin{equation}\label{STT}
S^{11}{_{kl}^{mn}}(\theta_{12})\,T^1{_{n\alpha}^{t\beta}}(\theta_1)\,
T^1{_{m\beta}^{s\gamma}}(\theta_2)=T^1{_{l\alpha}^{n\beta}}(\theta_2)\,
T^1{_{k\beta}^{m\gamma}}(\theta_1)\,S^{11}{_{mn}^{st}}(\theta_{12}),
\end{equation}
where the elements of the transmission matrix $T^1$ are infinite
dimensional. As noted in the previous section, the
transmission matrix elements have two types of label. The roman
labels stand for the soliton states $1,2,3$, while the greek labels
represent vectors in  the weight lattice. Because of the topological
charge conservation, the elements of the transmission matrix can be written as follows
\begin{equation}\label{Telements}
T^1{^{n\beta}_{i\alpha}}(\theta)=
t^1{_{i\alpha}^{n}}(\theta)\;\delta^{\beta-l_i+l_n}_{\alpha},\quad
i,n=1,2,3
\end{equation}
where $l_i,l_n$ are the weights \eqref{weights1}, which in the case
of $a_2$ are
\begin{equation}\label{weightsa2}
l_1=\frac{1}{3}(2\alpha_1+\alpha_2),\quad
l_2=-\frac{1}{3}(\alpha_1-\alpha_2),\quad
l_3=-\frac{1}{3}(\alpha_1+2\alpha_2).
\end{equation}
In the following discussion, indices referring to the first representation
will be omitted since there is no possibility of confusion;
for instance, the matrix $T^1$ will be indicated simply
by $T$, and so on. Using the ansatz \eqref{Telements} for the
transmission matrix and the $S$-matrix
\eqref{S11elements}, it is possible to find solutions to the
triangle equation \eqref{STT} up to an overall scalar function of
the rapidity. A classification of all possible solutions, and a
detailed explanation of the procedure adopted to obtain them, is
available in appendix \ref{appendixB}. Among all the solutions listed
\eqref{IIAIIBIIC} coincides with the $T$ matrix
\eqref{solitonsTmatrix} discovered already by analysing the bootstrap
procedure, as explained in the previous section. Notice that apart from
an overall scale this
solution contains eight parameters $t_{ij}$, satisfying the three relations
\eqref{tconstants}. However, using suitably designed
unitary transformations most of this freedom can be removed to leave just one essential
parameter.
To demonstrate this a slightly more general element of the transmission matrix $T$ will
be considered instead of the expression \eqref{Telements}, namely
\begin{equation}\label{Telementsmodified}
T^{n\beta p}_{i\alpha q}(\theta)= t_{i\alpha
p}^{n}(\theta)\;\delta^{\beta-l_i+l_n}_{\alpha}
\,\delta^{q+l_i+l_n}_{p}.
\end{equation}
This was the general expression considered using the functional
integral approach and the extra delta function does not alter the
solutions to the triangle equation
\eqref{Telements}.

\p It is convenient to split the argument into two steps.  Consider
the solution \eqref{IIAIIBIIC} and multiply it by
$(t_{11}t_{22}t_{33})^{-1/3}$. Next, conjugate the matrix using
the unitary transformation:
\begin{equation}\label{unitarytransdiag}
W_{\alpha p}^{\beta q}=(t_{11}^{-p\cdot l_2/2}t_{22}^{-p\cdot
l_1/2}t_{33}^{-p\cdot l_3/2})\,\delta_{\alpha}^{\beta}\delta_p^q,
\qquad |t_{11}|=|t_{22}|=|t_{33}|=1.
\end{equation}
After conjugation, the parametric part of the
solution \eqref{IIAIIBIIC} is modified and represented schematically as follows,
\begin{equation}\label{constantsjvl}
\left(%
\begin{array}{ccc}
  1 & t_{12}(t_{11}t_{22})^{-1/2}
  & t_{13}(t_{11}t_{33})^{-1/2} \\
  t_{21}(t_{11}t_{22})^{-1/2} & 1
   & t_{23}(t_{22}t_{33})^{-1/2} \\
 t_{31}(t_{11}t_{33})^{-1/2} &
 t_{32}(t_{22}t_{33})^{-1/2}& 1 \\
\end{array}%
\right)\equiv\left(%
\begin{array}{ccc}
  1 & \hat{t}_{12}
  & \hat{t}_{13}\\
  \hat{t}_{21} & 1
   & \hat{t}_{23}\\
 \hat{t}_{31} &
 \hat{t}_{32}& 1 \\
\end{array}%
\right).
\end{equation}
Next, conjugate using the unitary transformation,
\begin{equation}\label{unitarytransjvl}
V_{\alpha p}^{\beta q}=(\hat{t}_{21}^{-\alpha\cdot
\alpha_1/3}\hat{t}_{13}^{-\alpha
\cdot\alpha_0/3}\hat{t}_{32}^{-\alpha
\cdot\alpha_2/3})\,\delta_{\alpha}^{\beta}\delta_p^q,\qquad
|\hat{t}_{21}|=|\hat{t}_{13}|=|\hat{t}_{32}|=1,
\end{equation}
which, together with \eqref{tconstants}, transforms
\eqref{constantsjvl} to a matrix depending on a single parameter $t$,
which can be represented schematically by,
\begin{equation}
\left(%
\begin{array}{ccc}
  1 & t^{2/3} & t^{1/3} \\
  t^{1/3} & 1 & t^{2/3} \\
  t^{2/3} & t^{1/3} & 1 \\
\end{array}%
\right),
\end{equation}
\p with
$t\equiv(\hat{t}_{21}\hat{t}_{13}\hat{t}_{32})=(t_{21}t_{13}t_{32})/(t_{11}t_{22}t_{33})$.
Consequently, solutions \eqref{solitonsTmatrix} and
\eqref{antisolitonsTmatrix} become, respectively
\begin{eqnarray}\label{IIAIIBIICdelta}
&&T^1{^{n\beta}_{i\alpha}}(\theta)=g^1(\theta)
\, \left(
\begin{array}{ccc}
Q^{\alpha\cdot l_1}\,\delta_{\alpha}^{\beta}
  &\hat{x}^2\,\delta_{\alpha}^{\beta-\alpha_1}
  &\hat{x}\,Q^{-\alpha\cdot l_2}
  \, \delta_{\alpha}^{\beta+\alpha_0}\phantom{} \\
   \hat{x}\,Q^{-\alpha\cdot l_3}
  \,\delta_{\alpha}^{\beta+\alpha_1}
  &Q^{\alpha\cdot l_2}\,\delta_{\alpha}^{\beta}
  &\hat{x}^2\,\delta_{\alpha}^{\beta-\alpha_2}\\
  \hat{x}^2\,\delta_{\alpha}^{\beta-\alpha_0}
  &\hat{x}\,Q^{-\alpha\cdot l_1}
  \,\delta_{\alpha}^{\beta+\alpha_2}
  & Q^{\alpha\cdot l_3}
  \,\delta_{\alpha}^{\beta}\\
\end{array}
\right),\\
&&\nn\\
&&\nn\\
&&\ \ \ \ \ \ \ T^2{^{n\beta}_{i\alpha}}(\theta)=g^2(\theta)\label{IIAIIBIICantisolitonsdelta}
\,\left(
\begin{array}{ccc}
 Q^{-\alpha\cdot l_1}\,\delta_{\alpha}^{\beta}\phantom{}
  \,
  &\hat{x}\,\delta_{\alpha}^{\beta+\alpha_1}
  &0\\
  0
  &Q^{-\alpha\cdot l_2}\,\delta_{\alpha}^{\beta}\phantom{}
  &\hat{x}\,\delta_{\alpha}^{\beta+\alpha_2}\\
 \hat{x}\,\delta_{\alpha}^{\beta+\alpha_0}
  &0
  &Q^{-\alpha\cdot l_3}
  \,\delta_{\alpha}^{\beta}\phantom{} \\
\end{array}
\right),
\end{eqnarray}
where it has been convenient to set $$t\equiv e^{-3\gamma\Delta},$$
and
\begin{equation}\label{gtwo}
g^2(\theta)=g^1(\theta-i\pi/3)\;g^1(\theta+i\pi/3)
\;(1+\hat{x}^3), \quad \hat{x}=e^{\gamma(\theta-\Delta)}.
\end{equation}
Eventually, the constant $\Delta$ will be related to the Lagrangian parameter
$\sigma=e^{-\eta}$ introduced in \eqref{rootroot}.

\p
In the next section these  solutions to the triangle equations will be used as
suitable candidates for describing the jump-defect problem and an additional
constraint will be introduced to determine the
scalar function $g$ up to a CDD factor. Though the subsequent
analysis will rely on solution \eqref{IIAIIBIICdelta}, it could be
anticipated that this is not the only relevant solution of the triangle equation.
Evidence that it is the appropriate solution
 will be provided, as well as  reasons why
the functional integral approach selects solution \eqref{IIAIIBIIC}
among all the solutions presented in appendix \ref{appendixB}.

\section{The transmission matrix for the $a_2$ model: additional constraints}

\p Additional constraints are necessary to determine the
overall factor $g(\theta)$ in the solutions \eqref{IIAIIBIICdelta} and
\eqref{IIAIIBIICantisolitonsdelta}. For unitary theories these constraints
are based on unitarity and crossing properties of the S-matrix, although it
was found convenient in \cite{bczsg05} to use equivalent constraints based
on unitarity and `annihilation poles'.
The latter was found to be more suitable when analysing the sine-Gordon
system because it avoided having to relate scattering of solitons arriving at the defect
from the left to the scattering of solitons arriving at the defect from the right.
In the
present context, the theory is not unitary, the S-matrix is not a unitary matrix for
real rapidity, and it is not expected that the transmission matrix should
be unitary.
For this reason, the methods used previously to analyse the sine-Gordon model will
need to be adjusted slightly.

\p However, although the S-matrix is not unitary it is nevertheless natural to assume
that
\begin{equation}
S(-\theta)=[S(\theta)]^{-1}
\end{equation}
and therefore, a similar relation is also supposed to hold for the
transmission matrix \cite{Konik97}. The condition is
\begin{equation}\label{unitarity}
T^1{_{a\alpha}^{b\beta}}(\theta)\,\tilde{T}^1{_{b\beta}^{c\gamma}}(-\theta)
=\delta^{c}_{a}\delta^{\gamma}_{\alpha},
\end{equation}
where $\tilde{T}^1$ is the transmission matrix describing
the interaction between the defect and a soliton travelling from the
right to the left. In fact, since parity is violated explicitly in the
jump-defect problem, the matrix $\tilde{T}^1$ is expected to
differ from the matrix $T^1$ that describes solitons
travelling from left to right.
Indeed, the triangle equation
satisfied by  $\tilde{T}^1$  is
\begin{equation}\label{STTtilde}
S^{11}\,{_{lk}^{nm}}(\theta_{12})\,\tilde{T}^1{_{n\alpha}^{t\beta}}(\theta_1)\,
\tilde{T}^1{_{m\beta}^{s\gamma}}(\theta_2)=\,
\tilde{T}^1{_{l\alpha}^{n\beta}}(\theta_2)\,\tilde{T}^1{_{k\beta}^{m\gamma}}(\theta_1)\,
S^{11}\,{_{nm}^{ts}}(\theta_{12});
\end{equation}
and this differs slightly from the relation \eqref{STT} previously discussed.
Consequently, the solutions of these two triangle
equations are not the same. Nevertheless, $\tilde{T}^1(-\theta)$ is the
inverse of $T^1(\theta)$ and, therefore,
\begin{equation}\label{KIWtilde}
\tilde{T}^1{_{i\alpha}^{n\beta}}(-\theta)=\frac{1}{g^1(\theta)}\,\frac{1}{1-Q\hat{x}^3}
\,\left(
\begin{array}{ccc}
  Q^{-\alpha\cdot l_1}\,\delta_{\alpha}^{\beta}
  & 0
  & -Q\hat{x}\,\delta_{\alpha}^{\beta+\alpha_0}\\
  -Q\hat{x}\,\delta_{\alpha}^{\beta+\alpha_1}
  & Q^{-\alpha\cdot l_2}\,\delta_{\alpha}^{\beta}\phantom{}
  & 0\\
 0 &-Q\hat{x}\,\delta_{\alpha}^{\beta+\alpha_2}
  & Q^{-\alpha\cdot l_3}\,\delta_{\alpha}^{\beta}\phantom{} \\
\end{array}%
\right).
\end{equation}
It is worth pointing out  that requiring $T^1$ to have an inverse is already a constraint,
since not all solutions to the triangle equations will have this property.

\p
Crossing requires that \eqref{KIWtilde} should be closely related
to the transmission matrix $T^2$ for the antisoliton via the relation
\begin{equation}\label{crossing}
T^2{_{n\alpha}^{i\beta}}(\theta)=\tilde{T}^1{_{i\alpha}^{n\beta}}(i\pi-\theta).
\end{equation}
Comparing \eqref{KIWtilde} with \eqref{IIAIIBIICantisolitonsdelta} it is clear
\eqref{crossing} will be satisfied provided
\begin{equation}\label{firstconstraint}
g^1(\theta)\, g^1(\theta+i2\pi/3)\, g^1(\theta+i4\pi/3)\,\left(1-Q^3\hat{x}^3\right)\,
\left(1-Q\hat{x}^3\right)=1,
\end{equation}
which clearly constrains the overall factor $g^1(\theta)$. It is interesting that an
expression for $T^2$ emerges containing the zeroes remarked upon before, previously
generated by the bootstrap.

\p A minimal solution to \eqref{firstconstraint} is provided by setting
\begin{equation}\label{gfunction}
g^1(\theta)=\frac{f(\theta)}{(2\pi)^{2/3}\,\hat{x}} 
\end{equation}
\p with
\begin{equation}
f(\theta)=\Gamma[(1+\gamma)/2-z]\,\,\prod_{k=1}^{\infty}
\frac{\Gamma[(1+\gamma)/2+3k\gamma-z]\,
\Gamma[(1-\gamma)/2+(3k-2)\gamma+z]}
{\Gamma[(1-\gamma)/2+3k\gamma+z]\,
\Gamma[(1+\gamma)/2+(3k-1)\gamma-z]}\,,
\end{equation}
where $$z=\frac{i3\gamma(\theta-\Delta)}{2\pi}.$$
Using \eqref{gtwo} a matching expression can be found for $g^2(\theta)$:
\begin{equation}\label{gfunctiontwo}
g^2(\theta)=\frac{\Gamma[1/2 +\gamma-z]}{(2\pi)^{1/3}\, \hat{x}^{1/2}}\,
\prod_{k=1}^{\infty}
\frac{\Gamma[1/2+(3k+1)\gamma-z]\,
\Gamma[(1/2+(3k-2)\gamma+z]}
{\Gamma[(1/2+(3k-1)\gamma+z]\,
\Gamma[(1/2+(3k-1)\gamma-z]}\,.
\end{equation}
However, these expressions could be modified by multiplying $g^1(\theta)$
by any function $h(\theta)$ that
satisfies
$$h(\theta)\, h(\theta+i2\pi/3)\, h(\theta+i4\pi/3)=1,$$
and the ambiguity is not resolvable without comparing the results of the
algebraic manipulations with the outcome of
some alternative dynamical calculations. Unfortunately, such calculations are
beyond the scope of this article.

\p Since crossing has been used to constrain $g^1(\theta)$, and since the theory is not
unitary, there should be no further constraints. Previously, in \cite{bczsg05}, it
was found convenient to use the unitarity of the sine-Gordon model alongside the
annihilation poles. However, examining the annihilation poles in the present context merely
reproduces
\eqref{firstconstraint}. The `annihilation pole'
condition is provided by a  virtual process where a
particle and its antiparticle  annihilate to the vacuum and is described schematically
by the following expression
\begin{equation}\label{annihilationpole}
c_0{^{\bar{a}a}}\delta_{\alpha}^{\beta}=\sum_e\,
T^2\,{^{\bar{a}\beta}_{\bar{e}\gamma}}(\theta-i\pi/2)\,
T^1\,{^{a\gamma}_{e\alpha}}(\theta+i\pi/2)\,c_0{^{\bar{e}e}}.
\end{equation}
To perform the calculation it is necessary to determine the ratios of
the couplings appearing in \eqref{annihilationpole} by
examining the $S^{12}$ and $S^{21}$ matrix elements provided in
appendix \ref{appendixA}. When $\theta_{12}=i\pi$ the couplings are
\begin{eqnarray}
c{_{i\bar{\imath}}}^{0}\,c_{0}{^{\bar{k}k}}=c{_{\bar{\imath}i}}^{0}\,c_{0}{^{k\bar{k}}}=
\rho^{12}_0\,(q-q^{-1}), &&\quad i,k=1,2,3,\nn\\
c{_{j\bar{k}}}^{0}\,c_{0}{^{\bar{k}j}}=c{_{\bar{\jmath}\,k}}^{0}\,c_{0}{^{k\bar{\jmath}}}=0,
\ \ \ \ \ \ \ \ \ &&\quad j,k=1,2,3,\quad j\neq k,
\end{eqnarray}
where $\rho^{12}_0$ is the scalar function $\rho^{12}$ calculated
when $\theta_{12}=i\pi$. As a consequence, the coupling ratios
appearing in \eqref{annihilationpole} are all equal to one. Then,
using the transmission factors \eqref{IIAIIBIICdelta},
\eqref{IIAIIBIICantisolitonsdelta} in equation
\eqref{annihilationpole}, and setting  $a=1,2$ or $3$ , the `annihilation pole' condition
recovers \eqref{firstconstraint}.

\p Consider the pole occurring in the expression for $g^1(\theta)$ at $z=(1+\gamma)/2$,
or in terms of rapidity at
$$\theta_P =\Delta -\frac{i\pi}{3} -\frac{i\pi}{3\gamma}.$$
It is tempting to associate this pole with the complex rapidity at which the
classical delay diverges, namely \eqref{boundstate}, especially given that
$1/\gamma\rightarrow 0$ in the classical limit. That would then require the
identification
\begin{equation}\label{delta}
\Delta=\eta+\frac{i\pi}{2},
\end{equation}
at least in the limit $\beta\rightarrow 0$. With this identification,
the complex energy of the state associated with the pole at $\theta=\theta_P$
 is given by
\begin{equation}\label{poleenergy}
E=m_s\cosh\theta_P=m_s\cosh\eta\, \sin\left(\frac{\pi}{3}+\frac{\pi}{3\gamma}\right)
+im_s\sinh\eta \, \cos\left(\frac{\pi}{3}+\frac{\pi}{3\gamma}\right),
\end{equation}
and this enjoys a positive real part and  negative imaginary part provided
$$\frac{1}{2} <\gamma<2,$$
or, in terms of the coupling, $8\pi/3>\beta^2>4\pi/3$. Thus, it seems this
pole appears to indicate an unstable
state in the quantum theory that is completely disconnected from any
phenomenon in the classical model. This kind of feature did not appear
in the analysis of the sine-Gordon model.

\p
There are other reasons for making the identification \eqref{delta}, and reasons
related to breathers will be discussed in the next section.  However, comparison with
the sine-Gordon case already provides some additional motivation for
the choice. In fact, aligning with the notation used in the
present article, the transmission matrix for the sine-Gordon model
found in \cite{bczsg05} takes the form
\begin{equation}\label{}
T^{SG}{_{i\alpha}^{n\beta}}(\theta)=g^{SG}(\theta)\,\left(
  \begin{array}{cc}
    Q^{\alpha/2}\,\delta_{\alpha}^{\beta} &
    (-q)^{1/2}\,e^{\gamma(\theta-\eta)}\,\delta_{\alpha}^{\beta-2} \\
    (-q)^{1/2}\,e^{\gamma(\theta-\eta}\,\delta_{\alpha}^{\beta+2} &
    Q^{-\alpha/2}\,\delta_{\alpha}^{\beta}\\
  \end{array}
\right),\nn
\end{equation}
while for $a_2$, the transmission matrices found for solitons and
antisolitons, namely \eqref{IIAIIBIICdelta} and
\eqref{IIAIIBIICantisolitonsdelta}, with the choice \eqref{delta}
have remarkably similar elements since
\begin{equation}
\hat{x}=e^{\gamma(\theta-\Delta)}= (-q)^{-1/2}   e^{\gamma(\theta-\eta)}.
\end{equation}
Of course nothing can be said concerning the manner in which the classical
defect parameter $\eta$ might be renormalised, and in fact notation has been abused
slightly (though without leading to any misunderstandings)
by using the same symbol in two different contexts.

\p
At this stage, it is possible to compare the $a_2$ transmission matrices with the
available classical results, namely the delays \eqref{delay}
experienced by solitons or antisolitons travelling past the defect.
Classically, there is little difference in behaviour between the soliton and the
antisoliton. In either case the the defect causes a phase shift
varying between $0$ and $-2\pi/3$ for the
soliton (a=1), and between $0$ and $2\pi/3$ for an antisoliton
(a=2). Because of this shift, the topological charge
\eqref{topologicalcharges} of a soliton or antisoliton might change as
it passes the defect.
It was pointed out  that the topological charge of  a
soliton or antisoliton passing through
the defect could be converted to just one of the
adjacent topological charges. In particular, assuming $\theta>\eta$,
the argument of the delay \eqref{delayarg} will be negative for the soliton,
therefore its topological charge $l_i$ will change, if it changes at all, into
$l_{i+1}$ (with $(i+1)$ understood modulo $3$), while for the antisoliton the
argument of the delay will be positive and the topological charge $l_{\bar{\imath}}$
will change, if it changes, into $l_{\overline{\imath+1}}$
(with $(i+1)$ understood again modulo $3$). Inspecting
\eqref{IIAIIBIICantisolitonsdelta}, it can be seen that
the transmission matrix representing the behaviour of antisolitons provides a good
match to the
classical situation because of the presence of zeros in  expected
positions. On the other hand, the transmission matrix for  solitons
does not possess the expected zeros corresponding to the classical selection rule.
It appears that in the quantum context a
soliton passing through the defect may change into either of the
solitons adjacent to it; although the classically allowed transition remains
the most probable, the soliton can tunnel
to its classically forbidden neighbour. From this perspective, the defect
can act as a filter, which is intriguingly asymmetrical between solitons and
antisolitons. This kind of effect was not evident in the sine-Gordon model
since there the soliton and antisoliton belong to the same representation
when regarded from the perspective offered by the present context,
and transitions between the two are never forbidden, either classically or
quantum mechanically.

\section{Transmission factors for the lightest breathers}
\label{Breathers}

\p In order to collect additional evidence to support the idea that the
transmission matrix \eqref{IIAIIBIICdelta} describes $a_2$ solitons
interacting with a jump-defect, the transmission factors for the
lightest breathers will be calculated. Since the lightest breathers
correspond to the quantum Toda particles described by the
fundamental bulk fields appearing in the Lagrangian density, their
transmission factors can be compared perturbatively with  classical transmission
coefficients obtained by linearising  the defect conditions
\eqref{boundaryconditions}.

\p The breathers describe scalar bound states whose existence is
revealed by the following poles located in the forward channel of
the soliton-antisoliton scattering matrices (see appendix
\ref{appendixA})
\begin{equation}
\theta_k=i\pi \left(1-\frac{2\,k}{3\gamma}\right), \qquad
k=1,2,\dots [3\gamma/2], \quad k\in {\cal N},
\end{equation}
where the notation $[\mu]$ represents the largest integer less than
$\mu$. The masses of these bound states are
\begin{equation}
m_k=2\,M\sin\left(\frac{\pi k}{3\gamma}\right),
\end{equation}
where $M$ is the soliton mass \eqref{singlesolitonmasses}. The bootstrap
will be used to calculate the breather transmission factors and, for the lowest mass
breathers $(k=1)$, it states
\begin{equation}\label{lithestbreathers}
c_{1}{^{\bar{a}a}}\,T^{b_1}(\theta)\,\delta_{\alpha}^{\beta}
=\sum_{e}\,T^2\,{_{\bar{e}\alpha}^{\bar{a}\gamma}}\left(\theta-i(\pi/2-\pi/3\gamma)\right)\,
T^1\,{_{e\gamma}^{a\beta}}\left(\theta+i(\pi/2-\pi/3\gamma)\right)\,c_{1}{^{\bar{e}e}}.
\end{equation}
The ratios of the couplings can be calculated using the scattering
matrices provided in appendix \ref{appendixA}. For instance using
the matrix $S^{12}$, the couplings calculated at
$\theta_{12}=i\pi(1-2/3\gamma)$ satisfy
\begin{eqnarray}
c_{i\bar{\imath}}{^{1}}\,c_{1}{^{\bar{\imath}i}}&=&-\rho^{12}_1\,(q-q^{-1}),
\qquad\qquad c_{i\bar{m}}{^{1}}\,c_{1}{^{\bar{m}i}}=0, \qquad
i,m=1,2,3,\quad
m\neq i,\nn\\
c_{j\bar{\jmath}}{^{1}}\,c_{1}{^{\bar{l}l}}&=&\rho^{12}_1\,(q-q^{-1})\,e^{i\pi/3},\qquad\quad
c_{j\bar{\jmath}}{^{1}}\,c_{1}{^{\bar{k}k}}=\rho^{12}_1\,(q-q^{-1})\,e^{-i\pi/3},\nn\\
j&=&1,2,3 \quad k,l\neq j \qquad(k=j+1, l=j+2)\mod(3),\nn
\end{eqnarray}
where $\rho^{12}_1$ is the scalar function $\rho^{12}$ calculated
when $\theta_{12}=i\pi(1-2/3\gamma)$. Consequently, the coupling
ratios appearing in \eqref{lithestbreathers} are
\begin{equation}
\frac{c_{1}{^{\bar{3}3}}}{c_{1}{^{\bar{1}1}}}=\frac{c_{1}{^{\bar{1}1}}}{c_{1}{^{\bar{2}2}}}
=\frac{c_{1}{^{\bar{2}2}}}{c_{1}{^{\bar{3}3}}}=-e^{i\pi/3}, \qquad
\frac{c_{1}{^{3\bar{3}}}}{c_{1}{^{1\bar{1}}}}=\frac{c_{1}{^{1\bar{1}}}}{c_{1}{^{2\bar{2}}}}
=\frac{c_{1}{^{2\bar{2}}}}{c_{1}{^{3\bar{3}}}}=-e^{-i\pi/3}.
\end{equation}
Clearly, identical coupling ratios are obtained using the matrix $S^{21}$
instead of  $S^{12}$. Using  the transmission matrices
\eqref{IIAIIBIICdelta} and \eqref{IIAIIBIICantisolitonsdelta}, with
the scaling functions $g^1,\ g^2$ given by \eqref{gfunction}, \eqref{gfunctiontwo}, the transmission
factors for the lightest breathers are
\begin{equation}\label{breathersIIAIIBIICdelta}
T^{b_1}_1(\theta)=e^{-i\pi/3}\frac{\sinh\left(\frac{\theta-\eta}{2}-\frac{i\pi}{6}\right)}
{\sinh\left(\frac{\theta-\eta}{2}+\frac{i\pi}{6}\right)},\qquad
T^{b_1}_2(\theta)=e^{i\pi/3}\frac{\sinh\left(\frac{\theta-\eta}{2}+\frac{i2\pi}{3}\right)}
{\sinh\left(\frac{\theta-\eta}{2}+\frac{i\pi}{3}\right)}.
\end{equation}
Notice that, as was the case for sine-Gordon
\cite{bczsg05}, the transmission factors for the lightest breathers
appear to depend on the coupling constant $\beta$ only via the parameter $\eta$.
Nevertheless, one might expect, in the classical limit $\beta\rightarrow 0$, that the parameter
represented by $\eta$ appearing in
\eqref{breathersIIAIIBIICdelta} would tend to the classical Lagrangian parameter.
For this reason, as mentioned previously, the same notation has been used for this
parameter regardless of the context.

\p Consider now the classical problem of finding the transmission
coefficients for the linearized version of the jump-defect problem.
Following the procedure adopted in \cite{corrigan94} for the affine Toda field theory
restricted to a half line, the bulk fields $\phi$ and $\psi$ can be
expanded in power series in $\beta$ as follows,
$$\phi=\sum_{k=-1}^{\infty}\, \beta^k\,\phi^{(k)},\qquad
\psi=\sum_{k=-1}^{\infty}\, \beta^k\,\psi^{(k)}.$$ The fields
$\phi^{(0)}$ and $\psi^{(0)}$ represent the small coupling limit,
namely small perturbations around the background represented by the
fields $\phi^{(-1)}$ and $\psi^{(-1)}$. The field
$\phi^{(0)},\psi^{(0)}$ satisfy the linearized version of the
equations of motion and the defect conditions. Since the background
represents the ground state, it is supposed to have minimal energy
and to be time-independent. Any static configuration
\eqref{staticconfigurations}, as well as the choice
$(\phi^{(-1)},\psi^{(-1)})=(0,0)$, satisfy these requirements. Then,
the equations of motion and defect conditions for the fields
$\phi^{(0)},\psi^{(0)}$ become, respectively
\begin{eqnarray}\label{equationmotionlinearized}
\pt^2\phi^{(0)}
-\px^2\phi^{(0)}=-m^2\sum^{r}_{i=0}\alpha_{i}\alpha_i^T
\cdot\phi^{(0)}=-{\cal M}^2\phi^{(0)},\qquad &x<0,\nn\\
\pt^2\psi^{(0)}
-\px^2\psi^{(0)}=-m^2\sum^{r}_{i=0}\alpha_{i}\alpha_i^T
\cdot\psi^{(0)}=-{\cal M}^2\psi^{(0)},\qquad &x>0,
\end{eqnarray}
\begin{eqnarray}\label{defectconditionlinearized}
\partial_{x}\phi^{(0)}-E\pt \phi^{(0)}-D \pt
\psi^{(0)}+\frac{m\sigma}{4} \sum_{i=0}^{r}(\alpha_i \cdot D^T)
\left[ (\alpha_i \cdot D^T) \phi^{(0)}+
 (\alpha_i \cdot D) \psi^{(0)}\right]\nn\\
+\frac{m}{4\sigma} \sum_{i=0}^{r}(\alpha_i \cdot D)\left[( \alpha_i
\cdot D) \phi^{(0)}- (\alpha_i \cdot D) \psi^{(0)}\right]=0,
 \qquad x=0, \nn\\
\partial_{x}\psi^{(0)}-D^T\pt \phi^{(0)} +E\pt \psi^{(0)}
-\frac{m\sigma}{4} \sum_{i=0}^{r}(\alpha_i \cdot D )\left[(\alpha_i
\cdot D^T) \phi^{(0)}+ ( \alpha_i \cdot D)
\psi^{(0)}\right]\nn\\
+\frac{m}{4\sigma} \sum_{i=0}^{r}(\alpha_i \cdot D)\left[(\alpha_i
\cdot D) \phi^{(0)}- ( \alpha_i \cdot D) \psi^{(0)}\right]=0, \qquad
x=0, \label{boundary_conditions_multiSL2}
\end{eqnarray}
where ${\cal M}$ represents the mass matrix. A solution of the
equations of motion \eqref{equationmotionlinearized} is
\begin{eqnarray}\label{linearizedsolution}
\phi^{(0)}&=&\sum^{r}_{k=1}\rho_k\left(e^{i b_k x}+R_k e^{-i b_k
x}\right)e^{-i a_k t},\qquad x<0,\nn\\
\psi^{(0)}&=&\sum^{r}_{k=1}\rho_k T_k e^{i( b_k x-a_k t)},\qquad x>0.
\end{eqnarray}
The vector $\rho_k$, in order to satisfy the equations
\eqref{equationmotionlinearized}, has to be an eigenvector of the
matrix ${\cal M}$, namely
\begin{equation}\label{massmatrixequation}
{\cal M}^2\rho_k=(a_k^2-b_k^2)\rho_k=m_k^2\rho_k, \qquad
m_k=2\,m\sin\left(\frac{k\pi}{h}\right),\qquad k=1,\dots,r.
\end{equation}
By contrast with the half-line case discussed in
\cite{corrigan94}, the vectors $\rho_k$ do not diagonalise the
defect conditions \eqref{defectconditionlinearized}, because of the
presence of the matrix $D$. Therefore, explicit expressions
for these vectors are required in order to find the coefficients
$R_k,\ T_k$ appearing in \eqref{defectconditionlinearized}. Bearing in
mind the expression \eqref{todasoliton} for the soliton solutions,
the vectors $\rho_k$ can be written as follows
\begin{equation}\label{rhovectors}
\rho_k=-\sum_{l=0}^{r}\alpha_{l}\,\omega^{lk},\qquad \omega=e^{2\pi
i/h},\qquad k=1,\dots,r.
\end{equation}
It can be verified easily that these vectors satisfy
\eqref{massmatrixequation}.

\p Inserting \eqref{linearizedsolution} into the linearized defect
conditions \eqref{defectconditionlinearized}, two expressions
containing the unknown coefficients $R_k,\ T_k$ are obtained.
Multiplying them on the left hand side by $\rho_k^{\dagger}$, and
making use of \eqref{D} and \eqref{rhovectors}, leads to
\begin{eqnarray}
&&R_k[(-ib_k+ia_k+\sigma+1/\sigma)(2-\omega^{-k}-\omega^{k})-2ia_k(1-\omega^{k})]\nn\\
&&\ \ \ +T_k[2ia_k(1-\omega^{k})-(2-\omega^{-k}-\omega^{k})(\sigma \omega^{-k}+1/\sigma)]
+ib_k(2-\omega^{-k}-\omega^{k})\nn\\
&&\ \ \ \ \ +(\sigma+1/\sigma)(2-\omega^{-k}-\omega^{k})+ia_k(\omega^{k}-\omega^{-k})=0,
\qquad k=1,\dots,r,\nn\\
&&\nn\\
&&R_k[2ia_k(1-\omega^{-k})+(\sigma \omega^{-k}+1/\sigma)(2-\omega^{k}-\omega^{-k})]\nn\\
&&\ \ \ +T_k[ib_k(2-\omega^{-k}-\omega^{k})+ia_k(\omega^{-k}-\omega^{k})-(\sigma+1/\sigma)
(2-\omega^{k}-\omega^{-k})]\nn\\
&&\ \ \ \ \ +2ia_k(1-\omega^{-k})+(\sigma
\omega^{k}+1/\sigma)(2-\omega^{-k}-\omega^{k})=0,\qquad
k=1,\dots,r.\nn
\end{eqnarray}
After a little bit of algebra, and setting $a_k=m_k\cosh\theta$,
$b_k=m_k\sinh\theta$, the reflection and transmission coefficients
are found to be
\begin{eqnarray}\label{transmissioncoefficients}
R_k=0, \qquad T_k=\frac{ie^{-\theta}m_k-\sigma(1-\omega^{-k})} {i
e^{\theta}m_k+\sigma(1-\omega^{k})},\qquad k=1,\dots,r.
\end{eqnarray}
It can be easily verified for the $a_2$ affine Toda field theory, setting
$\sigma\equiv e^{-\eta}$,  that the transmission
coefficients \eqref{transmissioncoefficients} ($r=1,2$) coincide
with the expressions \eqref{breathersIIAIIBIICdelta} for the
transmission factors for the lightest breathers. Moreover, as expected,
the reflection factors turn out to be zero. This result,
given that no perturbative calculations are available to suggest
otherwise, provides a further justification for the choice made in \eqref{delta}
for the constant $\Delta$.

\section{On defects and solutions of the triangle equations}

\p The investigation of the triangle equations for the $a_2$ affine Toda field theory
reveals several possible candidates for the
transmission matrix describing the interaction between solitons and
the purely transmitting defects. In  previous sections, one of these
solutions has been chosen to be the `correct' matrix describing the
scattering between solitons and the jump-defect discussed,
classically, in section \ref{classicalATFT}. Some evidence to support
this  choice have been provided. However, for reasons of
completeness, some words are also due concerning the other solutions
listed in appendix B.

\p As  mentioned in section (\ref{classicalATFT}), the choice of the
`clockwise' cyclic permutation of the simple roots was arbitrary and
made simply in order to give a specific expression for the $D$
matrix. In fact, as was already pointed out,  the other possibility, using the
`anticlockwise' permutation, was chosen  in \cite{bcztoda}. It
turns out that if the alternative  choice had been adopted from the
start, the corresponding transmission matrix for solitons would have
been given by \eqref{KIW}, instead of \eqref{IIAIIBIIC}, and it
would have been the soliton transmission matrix that had the zeros
corresponding to  classical selection rules. Applying the bootstrap
procedure to
 \eqref{KIW} the resulting transmission matrix for the
antisolitons would be found to have no zero components.
As a consequence, for this alternative choice it is the matrix for the
solitons that mirrors the classical selection rules for the
delays of solitons and antisolitons passing through the defect.
Thus, reversing the sense of the permutation has the effect
of maintaining the asymmetry but interchanging the roles of solitons
and antisolitons.
 Similar
arguments  to those used to constrain the overall factor $g^1(\theta)$ can
be applied with the alternative choice of permutation leading to a
suitable overall scalar
function $\tilde{g}^1$. The transmission factors for the
lightest breathers can be calculated and, provided a suitable choice
for the single independent parameter appearing in the transmission
matrix is made, it can be verified they coincide with
\eqref{breathersIIAIIBIICdelta}, as was to be expected.

\p It should be noted that an alternative setting for the
jump-defect problem is also possible. Classically, the distinction
between the two settings turned out to be important in the process
of calculating of conserved charges \cite{bcztoda}. In fact,
according to which setting is chosen, only the even or odd spin
charges are conserved (apart from the spin $\pm 1$ charges that
correspond to energy and momentum). The defect conditions for the
alternative framework are
\begin{eqnarray}\label{boundaryconditionsnewsetting}
\partial_{x}\phi+E^T \pt \phi-D \pt \psi+\partial_\phi \cb=0
 \quad &x=0, \nn \\
\partial_{x}\psi-D^{T}\pt \phi -E^T\pt \psi
-\partial_\psi \cb=0 \quad&x=0.
\end{eqnarray}
with $D$ substituted by $-D$ in the defect potential
\eqref{affineTodadefectpotential}, and $D+D^T=-2$, $E=1+D^T$. An
explicit expression for the matrix $D$, provided a clockwise
permutation is chosen, is given by
\begin{equation}\label{D}
D=2\sum_{a=1}^r w_a\left(w_{a+1}-w_a\right)^T.
\end{equation}
In the quantum context the transmission matrix describing this
jump-defect framework is given by \eqref{VJL}. Alternatively,
choosing the anticlockwise permutation the correct transmission
matrix would be given by \eqref{IAIBIC}. All the computations
performed in this article can be repeated for this alternative
framework without any problems.

\p The remaining solutions of the triangle equations do not seem to
be relevant for the defect. Some of them fail to be invertible implying
they will never satisfy \eqref{unitarity}, and others fail to fit the
pattern implied by the functional arguments presented in section (\ref{functionalapproach}).

\section{Conclusion}

The purpose of this article has been to extend previous work devoted
to the sine-Gordon theory with a defect. During the analysis several
intriguing results have emerged. One of these, and perhaps the most
interesting, is the appearance of an unstable soliton-defect bound
state within a band of couplings that does not include a
neighbourhood of the classical limit. In a way, this is natural for
the $a_2$ model because  solitons cannot be absorbed by the defect
within the classical field theory; though a logical alternative
would have been a complete absence of unstable states. The next step
to take will be to examine the $a_n$ models in sufficient detail to
be able to determine the pattern of bound states accompanying
defect-soliton scattering. One of the first steps will be to analyse
a defect interacting with solitons whose topological charges are
described by weights in the six-dimensional representation
(corresponding to the centre spot in the Dynkin diagram) of $a_3$.
This is the first occasion where a classical model has missing
solutions (four of them, corresponding to four particular weights in
the ${\bf 6}$), and it will be interesting to see if there is a
mechanism to generate states corresponding to them in the quantum
theory.

\p Another intriguing feature is the manner by which the
quantum field theory with a defect chooses to implement the
classical selection rules governing the transitions between
different topological charges that are permitted by the defect. In
all cases (whether it be the choice of setting or permutation describing the defect in
the Lagrangian), there is an imbalance represented by the curious
asymmetry between the behaviour of solitons and the behaviour of
antisolitons represented typically by \eqref{IIAIIBIICdelta} and
\eqref{IIAIIBIICantisolitonsdelta}. Some difference between soliton
and antisoliton behaviour was to be expected owing to the explicit
breaking by the defect of parity and time-reversal but the way the
difference reveals itself is quite peculiar.

\p A further interesting fact
concerns the matrix $E$ (or equivalently $(1-D)$). This matrix is an ingredient of
the defect part of the classical Lagrangian and determined, in
the first place, by insisting upon classical integrability. However, in
section (\ref{functionalapproach}) it has been shown how it is
 alternatively  specified by examining the bootstrap in the functional
integral context.

\p It was pointed out in \cite{bczsg05} that it appears to be perfectly consistent
to allow several defects, or indeed to allow defects to move with independent velocities.
This part of sine-Gordon story has not been explored yet for the other
affine Toda theories and must be deferred for the moment.

\p One final remark. In most respects, members of the full set of affine Toda field models
share similar features, with such differences as there are attributable to their differing
root data. In gross
terms, a feature of one of them is a feature of the others. However, the classical analysis
of integrable defects has only revealed (so far) the possibility of defects within  the
$a_r$ series of models \cite{bcztoda}. On the other hand, all models in the imaginary coupling
regime have an $S$-matrix to describe the scattering of solitons, and one would expect
within each of these models a wide variety of infinite-dimensional solutions  to the triangle
equations. It remains to be seen if any of these solutions can be interpreted as soliton-defect
scattering though it would be surprising if such was not the case. Pushing the analysis in this
direction may shed some light on the existence (or otherwise) of a wider class of integrable
defect.

\eject
\vskip 1.0cm \noindent{\bf Acknowledgements} \vskip .25cm
\noindent One of us (CZ) is supported by a postdoctoral fellowship SPM 06-13 supplied by the
Centre National de la Recherche Scientifique (CNRS), and thanks the Japan Society
for the Promotion
of Science
for a Postdoctoral Fellowship when part of this work was performed; the other (EC)
thanks the Yukawa Institute for Theoretical
Physics, Kyoto University for hospitality on a number of occasions coinciding
with several stages of
this work. The work has also been supported in part by EUCLID (a
European Commission Research Training Network, contract number HPRN-CT-2002-00325).

\appendix
\section{S-matrices for the $a_2$ affine Toda field theory}
\label{appendixA}

\p For the $a_2$ affine Toda field theory, apart from the scattering matrix $S^{11}$
already described in section (\ref{Smatrix}), the matrices $S^{12}$
and $S^{21}$ are also used in the present article. Consider the
following triple product
\begin{eqnarray}
&&A^1_l(\theta_1)A^2_{\bar{k}}(\theta_2)\nn\\
&&\equiv A^1_l(\theta_1)[c_{\bar{k}}{^{ij}}\,A^1_i(\theta_2-i\pi/3)
A^1_j(\theta_2+i\pi/3)+c_{\bar{k}}{^{ji}}\,A^1_j(\theta_2-i\pi/3)
A^1_i(\theta_2+i\pi/3)],\qquad l^2_{k}=l_i+l_j,\nn
\end{eqnarray}
where $l^2_k\equiv l_{\bar{k}}= -l_k$ ($k=1,2,3$) and the value of
the couplings is given in \eqref{solitons2}. Then, making use of
\eqref{S11elements}, the non-zero components of the matrix $S^{12}$
are given by
\begin{eqnarray}
A^1_{j}(\theta_1)A^2_{\bar{\jmath}}(\theta_2)&=&\sum_{k=1}^{3}
S^{12}\,{^{\bar{k}\,k}_{j\,\bar{\jmath}}}\,(\theta_{12})A^2_{\bar{k}}(\theta_2)
A^1_{k}(\theta_1),\nn\\
\nn\\
A^1_{j}(\theta_1)A^2_{\bar{k}}(\theta_2)&=&
S^{12}\,{^{\bar{k}\,j}_{j\,\bar{k}}}\,(\theta_{12})A^2_{\bar{k}}(\theta_2)A^1_{j}(\theta_1),
\quad j\neq k,
\end{eqnarray}
with 
\begin{eqnarray}\label{S12elements}
S^{12}\,{^{\bar{\jmath}\,j}_{j\,\bar{\jmath}}}\,
(\theta_{12})&=&\phantom{-}\rho^{12}(\theta_{12})\,
\left(x_{12}(-q)^{1/2}-x_{12}^{-1}(-q)^{-1/2}\right),\nn\\
\nn\\
S^{12}\,{^{\bar{k}\,k}_{j\,\bar{\jmath}}}\,(\theta_{12})&=&\phantom{-}\rho^{12}(\theta_{12})
\,\left(q-q^{-1}\right)
\left\{%
\begin{array}{ll}
x_{12}^{\phantom{-}1/3}(-q)^{\phantom{-}1/2}, \ \ k=j-1\  \mod(3) \\
\\
x_{12}^{-1/3}(-q)^{-1/2}, \ \ k=j+1\ \mod(3)\\
\end{array}%
\right.\nn\\
\nn\\
S^{12}\,{^{\bar{k}\,j}_{j\,\bar{k}}}\,(\theta_{12})&=&-
\rho^{12}(\theta_{12})\,\left(x_{12}(-q)^{3/2}-x_{12}^{-1}(-q)^{-3/2}\right),
\quad j\neq k,\label{S12multiplier}
\end{eqnarray}
where the scalar mulitplier $\rho^{12}$ is
\begin{equation}
\rho^{12}(\theta_{12})=
\left(x_{12}(-q)^{-1/2}-x_{12}^{-1}(-q)^{1/2}\right)
\,\rho^{11}(\theta_{12}+\pi i/3) \, \rho^{11}(\theta_{12}-\pi i/3).\nn
\end{equation}
The non-zero elements of the matrix $S^{21}$ are  equal to the
elements described in \eqref{S12elements} for the matrix $S^{12}$
with $\rho^{12}=\rho^{21}$, except for the
$S^{21}\,{^{k\,\bar{k}}_{\bar{\jmath}\,j}}$ elements, for which the
index $k=j-1$ has to be replaced by $k=j+1$, and vice versa. Such a
small difference turns out to be relevant in the calculation of the
transmission factors for the lightest breathers performed in section
(\ref{Breathers}). A detailed investigation of the $a_2$ affine Toda
field theory, including bound states and scattering processes, can
be found in \cite{Gandenberger95}.

\section{Solutions of the triangle equations for the $a_2$ model}
\label{appendixB}

\p A classification of the possible solutions of the Yang-Baxter
equation for purely transmitting defects \eqref{STT} will be
provided in this appendix. As already explained in section
(\ref{section5}), because of the topological charge conservation, the
ansatz for the elements of the transmission matrix is supplied by
\eqref{Telements}. For the analysis performed in this appendix, it
is useful to assign a different letter to each entries of the
transmission matrix to avoid the use of many indexes. The notation
chosen is the following
\begin{equation}T^{j\beta}_{i\alpha}(\theta)\equiv \left(
\begin{array}{ccc}
  A_{\alpha}^{\beta}(\theta) & K_{\alpha}^{\beta}(\theta)
  & V_{\alpha}^{\beta}(\theta)\\
  J_{\alpha}^{\beta}(\theta)&
B_{\alpha}^{\beta}(\theta)
  & I_{\alpha}^{\beta}(\theta) \\
  W_{\alpha}^{\beta}(\theta)& L_{\alpha}^{\beta}(\theta)
  & C_{\alpha}^{\beta}(\theta)\\
\end{array}
\right)=\left(
\begin{array}{ccc}
  a_{\alpha}(\theta)\,\delta_{\alpha}^{\beta}\phantom{aa} & k_{\alpha}(\theta)\,
\delta_{\alpha}^{\beta-\alpha}\,\phantom{\delta^{\alpha}}
  & v_{\alpha}(\theta)\, \delta_{\alpha}^{\beta+\alpha_0} \\
  j_{\alpha}(\theta)\,
\delta_{\alpha}^{\beta+\alpha_1} &
b_{\alpha}(\theta)\,\delta_{\alpha}^{\beta}\,\phantom{\delta^{\alpha}}\phantom{aa}
  & i_{\alpha}(\theta)\,\delta_{\alpha}^{\beta-\alpha_2} \\
  w_{\alpha}(\theta)\,
\delta_{\alpha}^{\beta-\alpha_0} & l_{\alpha}(\theta)\,
\delta_{\alpha}^{\beta+\alpha_2}\phantom{a}
  & c_{\alpha}(\theta)\,\delta_{\alpha}^{\beta}\,\phantom{\delta^{\alpha}}\phantom{}\nn\\
\end{array}
\right).
\end{equation}
In addition, the following short notation will be adopted
\begin{equation}
(q\,x_{12}-q^{-1}\,x_{12}^{-1})\equiv a, \qquad
(x_{12}-x_{12}^{-1})\equiv b, \qquad (q-q^{-1})\equiv c,
\end{equation}
with
$$x_{12}=\frac{x_1}{x_2},\qquad x_j=e^{3\gamma\theta/2}, \qquad
q=-e^{-i\pi\gamma}.$$ As a starting assumption, all entries of the
transmission matrix are supposed to be different from zero.
Expression \eqref{STT} provides several relations involving the
elements of the $T$ matrix, which can be gathered into five groups. For
each group of relations, examples will be provided. The notation
$A_1$, $A_2$ etc. will be used to indicate the entries
$A(\theta_1)$, $A(\theta_2)$, respectively.
\begin{itemize}
    \item Group 1
    \begin{equation}\label{group1}
A_1A_2=A_2A_1,
\end{equation}
\end{itemize}
and eight more equations, one for each entry of the $T$ matrix. For the
diagonal entries, this kind of relation is automatically satisfied,
while for the other entries they state that the ratios
\begin{eqnarray}\label{constraint1}
\frac{k_{\alpha+\alpha_1}}{k_{\alpha}}, \qquad
\frac{j_{\alpha+\alpha_1}}{j_{\alpha}}, \qquad
\frac{v_{\alpha+\alpha_0}}{v_{\alpha}},\qquad
\frac{w_{\alpha+\alpha_0}}{w_{\alpha}}, \qquad
\frac{i_{\alpha+\alpha_2}}{i_{\alpha}}, \qquad
\frac{l_{\alpha+\alpha_2}}{l_{\alpha}},
\end{eqnarray}
are independent of rapidity.
\begin{itemize}
    \item Group 2
\begin{eqnarray}\label{group2}
b(A_1B_2-B_2A_1)&=&c(J_2K_1x_{12}^{-1/3}-J_1K_2x_{12}^{1/3}),\nn\\
b(B_1A_2-A_2B_1)&=&c(K_2J_1x_{12}^{1/3}-K_1J_2x_{12}^{-1/3}),\nn\\
c\,x_{12}^{1/3}(B_1A_2-B_2A_1)&=&b(J_2K_1-K_1J_2),\nn\\
c\,x_{12}^{-1/3}(A_1B_2-A_2B_1)&=&b(K_2J_1-J_1K_2),
\end{eqnarray}
\end{itemize}
and another two  similar series of four relations involving  the
elements $A,C,W,V$ and $B,C,I,L$, respectively. The first two
expressions in \eqref{group2} force the ratio
$(x^{-2/3}\;k_{\alpha})/j_{\alpha}$ to be independent of rapidity.
The remaining two expressions are not independent since their sum
turns out to be zero. Therefore, only one expression in
\eqref{group2} has still to be analyzed. Similar conclusions can be
drawn from the other two series of relations. In the end the constraints state that
\begin{equation}\label{constraint2}
\frac{k_{\alpha}}{j_{\alpha}}x^{-2/3}, \qquad
\frac{w_{\alpha}}{v_{\alpha}}x^{-2/3}, \qquad
\frac{i_{\alpha}}{l_{\alpha}}x^{-2/3},
\end{equation}
are independent of rapidity and three expressions remain to be analyzed. The latter
will be kept on one side to be discussed at the end of this section.
\begin{itemize}
\item Group 3
\begin{eqnarray}\label{group3a}
aA_1K_2-bK_2A_1=c\,x_{12}^{-1/3}A_2K_1,&\qquad
&aK_1A_2-bA_2K_1=c\,x_{12}^{1/3}K_2A_1,\nn\\
aA_2J_1-bJ_1A_2=c\,x_{12}^{-1/3}A_1J_2,&\qquad
&aJ_2A_1-bA_1J_2=c\,x_{12}^{1/3}J_1A_2,\\
\nn\\
aA_1V_2-bV_2A_1=c\,x_{12}^{1/3}A_2V_1,&\qquad
&aV_1A_2-bA_2V_1=c\,x_{12}^{-1/3}V_2A_1,\nn\\
aA_2W_1-bW_1A_2=c\,x_{12}^{1/3}A_1W_2,&\qquad
&aW_2A_1-bA_1W_2=c\,x_{12}^{-1/3}W_1A_2\label{group3b}.
\end{eqnarray}
\end{itemize}
The relations \eqref{group3a} are satisfied if
$a_{\alpha+\alpha_1}/a_{\alpha}=q$ and the two ratios
$a_{\alpha}\,x^{-2/3}/k_{\alpha}$ and
$a_{\alpha}\,x^{-4/3}/j_{\alpha}$ are independent of rapidity; or, if
$a_{\alpha+\alpha_1}/a_{\alpha}=1/q$ and the ratios
$a_{\alpha}\,x^{4/3}/k_{\alpha}$, $a_{\alpha}\,x^{2/3}/j_{\alpha}$
are independent of rapidity. In short, these solutions are
summarized as follows
\begin{eqnarray}\label{constraint3A1}
\frac{a_{\alpha+\alpha_1}}{a_{\alpha}}=q, \quad
\frac{a_{\alpha}}{k_{\alpha}}x^{-2/3}, \quad
\frac{a_{\alpha}}{j_{\alpha}}x^{-4/3}\quad \mbox{or}\quad
\frac{a_{\alpha+\alpha_1}}{a_{\alpha}}=\frac{1}{q}, \quad
\frac{a_{\alpha}}{k_{\alpha}}x^{4/3}, \quad
\frac{a_{\alpha}}{j_{\alpha}}x^{2/3}.
\end{eqnarray}
Similarly, for the expressions \eqref{group3b} the solutions are
\begin{eqnarray}\label{constraint3A2}
\frac{a_{\alpha+\alpha_0}}{a_{\alpha}}=\frac{1}{q}, \quad
\frac{a_{\alpha}}{v_{\alpha}}x^{-4/3}, \quad
\frac{a_{\alpha}}{w_{\alpha}}x^{-2/3} \quad
\mbox{or}\quad\frac{a_{\alpha+\alpha_0}}{a_{\alpha}}=q, \quad
\frac{a_{\alpha}}{v_{\alpha}}x^{2/3}, \quad
\frac{a_{\alpha}}{w_{\alpha}}x^{4/3}.
\end{eqnarray}
In group 3 there are further relations, among which there are eight involving the
elements $B$ and $K,J,L,I$ and another eight involving $C$ and $V,W,L,I$.
Using the same notation as \eqref{constraint3A1} and
\eqref{constraint3A2}, the constraints expressed by the other
relations of  this group, which are not listed here, are
\begin{eqnarray}\label{constraint3B1}
\frac{b_{\alpha+\alpha_1}}{b_{\alpha}}=q, \quad
\frac{b_{\alpha}}{k_{\alpha}}x^{4/3}, \quad
\frac{b_{\alpha}}{j_{\alpha}}x^{2/3}&\quad \mbox{or}\quad
&\frac{b_{\alpha+\alpha_1}}{b_{\alpha}}=\frac{1}{q}, \quad
\frac{b_{\alpha}}{k_{\alpha}}x^{-2/3}, \quad
\frac{b_{\alpha}}{j_{\alpha}}x^{-4/3}\\
\frac{b_{\alpha+\alpha_2}}{b_{\alpha}}=q, \quad
\frac{b_{\alpha}}{i_{\alpha}}x^{-2/3}, \quad
\frac{b_{\alpha}}{l_{\alpha}}x^{-4/3}&\quad \mbox{or}\quad
&\frac{b_{\alpha+\alpha_2}}{b_{\alpha}}=\frac{1}{q}, \quad
\frac{b_{\alpha}}{i_{\alpha}}x^{4/3}, \quad
\frac{b_{\alpha}}{l_{\alpha}}x^{2/3}\label{constraint3B2}
\end{eqnarray}
and
\begin{eqnarray}\label{constraint3C1}
\frac{c_{\alpha+\alpha_0}}{c_{\alpha}}=q, \quad
\frac{c_{\alpha}}{v_{\alpha}}x^{-4/3}, \quad
\frac{c_{\alpha}}{w_{\alpha}}x^{-2/3}&\quad \mbox{or}\quad
&\frac{c_{\alpha+\alpha_0}}{c_{\alpha}}=\frac{1}{q}, \quad
\frac{c_{\alpha}}{v_{\alpha}}x^{2/3}, \quad
\frac{c_{\alpha}}{w_{\alpha}}x^{4/3}\\
\frac{c_{\alpha+\alpha_2}}{c_{\alpha}}=q, \quad
\frac{c_{\alpha}}{i_{\alpha}}x^{4/3}, \quad
\frac{c_{\alpha}}{l_{\alpha}}x^{2/3}&\quad \mbox{or}\quad
&\frac{c_{\alpha+\alpha_2}}{c_{\alpha}}=\frac{1}{q}, \quad
\frac{c_{\alpha}}{i_{\alpha}}x^{-2/3}, \quad
\frac{c_{\alpha}}{l_{\alpha}}x^{-4/3}\label{constraint3C2}.
\end{eqnarray}
Clearly, the results of the group 3 can be gathered in turn into three
subgroups, which will be called $3A$, $3B$ and $3C$ because of the
fact that their relations incorporate the diagonal entries $A$, $B$
or $C$ of the $T$ matrix. Note that each of these subgroups provides
four possible different solutions, according to the combination chosen.
For instance, for the subgroup $3A$ it is possible to choose the
first expression in \eqref{constraint3A1} together with the first
expression in \eqref{constraint3A2}, or the second one. These are
already two different combinations. Similarly, starting with the
second expression in \eqref{constraint3A1}. Note also that each
subgroup provides a complete understanding of the dependence of the
diagonal elements of the $T$ matrix with respect to the simple
roots, and therefore with respect to a general vector
$\alpha=m\,\alpha_1+n\,\alpha_2$. In fact, it is possible to
conclude that the ratios $a_{\alpha+\alpha_2}/a_{\alpha}$,
$b_{\alpha+\alpha_0}/b_{\alpha}$ and
$c_{\alpha+\alpha_1}/c_{\alpha}$ can only be equal to 1 or $1/q^2$
or $q^2$. This piece of information will be relevant for the
analysis of the next group of equations.
\begin{itemize}
\item Group 4
\end{itemize}
$$b(A_1I_2-I_2A_1)=c\,x_{12}^{1/3}(J_2V_1-J_1V_2);\
b(I_1A_2-A_2I_1)=c\,x_{12}^{-1/3}(V_2J_1-V_1J_2),$$
$$b(J_1V_2-V_2J_1)=c(x_{12}^{1/3}A_2I_1-x_{12}^{-1/3}A_1I_2);\
b(V_1J_2-J_2V_1)=c(x_{12}^{-1/3}I_2A_1-x_{12}^{1/3}I_1A_2),$$
$$b(L_1A_2-A_2L_1)=c\,x_{12}^{1/3}(K_2W_1-K_1W_2);\
b(A_1L_2-L_2A_1)=c\,x_{12}^{-1/3}(W_2K_1-W_1K_2),$$
$$b(K_1W_2-W_2K_1)=c(x_{12}^{1/3}L_2A_1-x_{12}^{-1/3}L_1A_2);\
b(W_1K_2-K_2W_1)=c(x_{12}^{-1/3}A_2L_1-x_{12}^{1/3}A_1L_2),$$

\p with sixteen other similar relations, eight involving the element $B$ together
with all the off diagonal elements of the $T$ matrix, and eight involving
the $C$ element together with all the off diagonal entries. The
first constraint provided by all these equations is the following
\begin{equation}\label{constraint4}
\frac{a_{\alpha+\alpha_2}}{a_{\alpha}}=\frac{b_{\alpha+\alpha_0}}{b_{\alpha}}
=\frac{c_{\alpha+\alpha_1}}{c_{\alpha}}=1.
\end{equation}
In other words, the other possibilities mentioned previously are not
permitted, since they contradict \eqref{constraint2}. This observation
allows a reduction in the number of possible combinations of solutions in
each group $3A$, $3B$ and $3C$ from four to two. At this stage, for
pursuing the analysis of the triangular relations, it is useful
to adopt the following notation for the ratios independent of
rapidity appearing in \eqref{constraint3A1}-\eqref{constraint3C2}:
\begin{equation}\label{notationforh}
\frac{a_\alpha}{p_\alpha} x^{\pm\epsilon_p/3}=\frac{1}{h_{a
p}(\alpha)}\frac{t_a}{t_p},\,\quad \frac{b_\alpha}{p_\alpha}
x^{\pm\epsilon_p/3}=\frac{1}{h_{b p}(\alpha)}\frac{t_b}{t_p},\,\quad
\frac{c_\alpha}{p_\alpha} x^{\pm\epsilon_p/3}=\frac{1}{h_{c
p}(\alpha)}\frac{t_c}{t_p},\quad \epsilon_p=2,4,
\end{equation}
where $h_{kj}$ are exponential functions, $t_k$ are constants and
$p_\alpha$ stands for one of the off-diagonal entries of the $T$
matrix appearing in expressions
\eqref{constraint3A1}-\eqref{constraint3C2}. With this notation, the
constraints provided by the relations in  group 4 can be summarized
as follows.

\p Consider the relations in the subgroup $3A$, namely
\eqref{constraint3A1} and \eqref{constraint3A2}. If the combination
\begin{equation}\label{ratiosa1}
\frac{a_{\alpha+\alpha_1}}{a_{\alpha}}=q,\qquad
\frac{a_{\alpha+\alpha_0}}{a_{\alpha}}=\frac{1}{q},
\end{equation} \p holds,
then
\begin{equation}
\frac{h_{aj}(\alpha_0)}{h_{av}(\alpha_1)}=q^2, \qquad
\frac{h_{ak}(\alpha_0)}{h_{aw}(\alpha_1)}=1,\qquad
h_{ak}(\alpha)h_{aw}(\alpha+\alpha_1)t_k t_w=h_{al}(\alpha)t_l
t_a.\nn
\end{equation}
On the other hand, if
\begin{equation}\label{ratiosa1}
\frac{a_{\alpha+\alpha_1}}{a_{\alpha}}=\frac{1}{q},\qquad
\frac{a_{\alpha+\alpha_0}}{a_{\alpha}}=q,
\end{equation}
then
\begin{equation}
\frac{h_{aj}(\alpha_0)} {h_{av}(\alpha_1)}=1, \qquad
\frac{h_{ak}(\alpha_0)} {h_{aw}(\alpha_1)}=\frac{1}{q^2},\qquad
h_{aj}(\alpha)h_{av}(\alpha-\alpha_1)t_j t_v=h_{ai}(\alpha)t_i
t_a.\nn
\end{equation}
Similarly, consider the relations in the subgroup $3B$. If
\begin{equation}\label{ratiosa1}
\frac{b_{\alpha+\alpha_1}}{b_{\alpha}}=\frac{1}{q},\qquad
\frac{b_{\alpha+\alpha_2}}{b_{\alpha}}=q,
\end{equation}
then
\begin{equation}
\frac{h_{bl}(\alpha_1)} {h_{bj}(\alpha_2)}=q^2, \qquad
\frac{h_{bi}(\alpha_1)} {h_{bk}(\alpha_2)}=1,\qquad
h_{bi}(\alpha)h_{bk}(\alpha+\alpha_2)t_i t_k=h_{bv}(\alpha)t_v
t_b.\nn
\end{equation}
On the other hand, if
\begin{equation}\label{ratiosa1}
\frac{b_{\alpha+\alpha_1}}{b_{\alpha}}=q,\qquad
\frac{b_{\alpha+\alpha_2}}{b_{\alpha}}=\frac{1}{q},
\end{equation}
then
\begin{equation}
\frac{h_{bl}(\alpha_1)} {h_{bj}(\alpha_2)}=1, \qquad
\frac{h_{bi}(\alpha_1)} {h_{bk}(\alpha_2)}=\frac{1}{q^2},\qquad
h_{bl}(\alpha)h_{bj}(\alpha-\alpha_2)t_l t_j=h_{bw}(\alpha)t_w
t_b\nn.
\end{equation}
Finally, looking at the relations in the subgroup $3C$. If
\begin{equation}\label{ratiosa1}
\frac{c_{\alpha+\alpha_0}}{c_{\alpha}}=q,\qquad
\frac{c_{\alpha+\alpha_2}}{c_{\alpha}}=\frac{1}{q},
\end{equation}
then
\begin{equation}
\frac{h_{cv}(\alpha_2)} {h_{cl}(\alpha_0)}=q^2, \qquad
\frac{h_{cw}(\alpha_2)} {h_{ci}(\alpha_0)}=1,\qquad
h_{ci}(\alpha)h_{cw}(\alpha+\alpha_2)t_i t_w=h_{cj}(\alpha)t_j
t_c.\nn
\end{equation}
Instead, if
\begin{equation}\label{ratiosa1}
\frac{c_{\alpha+\alpha_0}}{c_{\alpha}}=\frac{1}{q},\qquad
\frac{c_{\alpha+\alpha_2}}{c_{\alpha}}=q,
\end{equation}
then
\begin{equation}
\frac{h_{cv}(\alpha_2)} {h_{cl}(\alpha_0)}=1, \qquad
\frac{h_{cw}(\alpha_2)} {h_{ci}(\alpha_0)}=\frac{1}{q^2},\qquad
h_{cl}(\alpha)h_{cv}(\alpha-\alpha_2)t_l t_v=h_{ck}(\alpha)t_k
t_c.\nn
\end{equation}
Finally, the last group of equations is
\begin{itemize}
\item Group 5
\begin{eqnarray}\label{group5}
aV_1K_2-bK_2V_1=c\,x_{12}^{1/3}V_2K_1, \quad
aK_1V_2-bV_2K_1=c\,x_{12}^{-1/3}K_2V_1,\nn\\
aW_2J_1-bJ_1W_2=c\,x_{12}^{1/3}W_1J_2, \quad
aJ_2W_1-bW_1J_2=c\,x_{12}^{-1/3}J_1W_2,\nn
\end{eqnarray}
\end{itemize}
with eight further relations. None of these involve the diagonal terms of
the transmission matrix. The constraints provided by them can be
summarized as follows
\begin{equation}
\frac{h_{ak}(-\alpha_0)} {h_{av}(\alpha_1)}= \frac{h_{aj}(\alpha_0)}
{h_{aw}(-\alpha_1)}=q,\quad \frac{h_{bl}(\alpha_1)}
{h_{bk}(-\alpha_2)}= \frac{h_{bi}(-\alpha_1)}
{h_{bj}(\alpha_2)}=q,\quad \frac{h_{cw}(-\alpha_2)}
{h_{cl}(\alpha_0)}= \frac{h_{cv}(\alpha_2)}
{h_{ci}(-\alpha_0)}=q.\nn
\end{equation}
All these constraints taken together allow  eight families of
possible solutions that can be written down explicitly. Firstly,
note that the functions of the type $h_{ap}(\alpha)$ can be split
into the ratios $h_{p}(\alpha)/h_{a}(\alpha)$ (see
\eqref{notationforh}), where each
$$h_p(\alpha)=q^{\alpha(m_p\,\alpha_1+n_p\,\alpha_2)}$$ with $m_p$ and
$n_p$ constants. Bearing this in mind, and taking into account that
the values of the constants $m_p$ and $n_p$ for the diagonal entries
are already known, it is possible to simplify the relations in
groups 4 and 5 involving the functions $h_{ap}(\alpha)$, and
determine the constants $m_p$ and $n_p$ for the other entries. In
order to be as clear  as possible, and to avoid the use of a
heavy notation, it is sufficient to write down explicitly as an example only
one solution for each family. The choice of the explicit solutions, which
may be called
`minimal', is motivated by the fact that the solutions relevant for
the defect problem lie within this group. Despite that, an
example of a complete family of solutions will be provided later,
and the use of the term `minimal' should become clearer.

\p Rewriting the constants $t_p$ as $t_{ij}$, indicating their
positions in the transmission matrix (for instance, $t_k=t_{12}$),
the eight `minimal' solutions found - one for each family - are
$(q^{-1}\equiv Q)$:

\begin{itemize}
\item $t_{21}\,t_{13}=t_{23},\qquad
\frac{t_{32}\,t_{21}}{t_{22}}=t_{31},\qquad
\frac{t_{32}\,t_{13}}{t_{33}}=t_{12},$

\begin{equation}\label{IIAIIBIIC}
\left(%
\begin{array}{ccc}
Q^{\alpha \cdot l_1}\,\delta_{\alpha}^{\beta}\phantom{}
  \,
  & t_{12}\,x^{4/3}\,\delta_{\alpha}^{\beta-\alpha_1}\,\phantom{}
  & t_{13}\,x^{2/3}\,Q^{-\alpha \cdot l_2}
  \, \delta_{\alpha}^{\beta+\alpha_0}\phantom{} \\
  t_{21}\,x^{2/3}\,Q^{-\alpha \cdot l_3}
  \,\delta_{\alpha}^{\beta+\alpha_1}\phantom{}
  &
  t_{22}\,Q^{\alpha \cdot l_2}\,\delta_{\alpha}^{\beta}\phantom{}
  & t_{23}\,x^{4/3}\,\delta_{\alpha}^{\beta-\alpha_2}\phantom{} \\
 t_{31}\,x^{4/3}\,\delta_{\alpha}^{\beta-\alpha_0}\phantom{}
  & t_{32}\,x^{2/3}\,Q^{-\alpha\cdot l_1}
  \,\delta_{\alpha}^{\beta+\alpha_2}\phantom{}
  & t_{33}\,Q^{\alpha\cdot l_3}
  \,\delta_{\alpha}^{\beta}\phantom{} \\
\end{array}
\right)
\end{equation}

\vspace{.3cm}
\item $t_{12}\,t_{31}=t_{32},\qquad
\frac{t_{23}\,t_{21}}{t_{22}}=t_{13},\qquad
\frac{t_{23}\,t_{31}}{t_{33}}=t_{21},$

\begin{equation}\label{IAIBIC}
\left(%
\begin{array}{ccc}
Q^{-\alpha\cdot l_1}\,\delta_{\alpha}^{\beta}\phantom{}
  \,
  &
  t_{12}\,x^{-2/3}\,Q^{\alpha\cdot l_3}\;\delta_{\alpha}^{\beta-\alpha_1}
  & t_{13}\,x^{-4/3}
  \, \delta_{\alpha}^{\beta+\alpha_0}\phantom{} \\
  t_{21}\,x^{-4/3}
  \,\delta_{\alpha}^{\beta+\alpha_1}\phantom{}
  &
  t_{22}\,Q^{-\alpha\cdot l_2}\,
  \delta_{\alpha}^{\beta}\phantom{}
  & t_{23}\,x^{-2/3}\,Q^{\alpha\cdot l_1}\,
  \delta_{\alpha}^{\beta-\alpha_2} \\
t_{31}\,x^{-2/3}\,Q^{\alpha\cdot l_2}\,
 \delta_{\alpha}^{\beta-\alpha_0}
  & t_{32}\,x^{-4/3}
  \,\delta_{\alpha}^{\beta+\alpha_2}\phantom{}
  & t_{33}\,Q^{-\alpha\cdot l_3}
  \,\delta_{\alpha}^{\beta}\phantom{} \\
\end{array}
\right)
\end{equation}

\end{itemize}

\p These two solutions are the solutions used earlier as the soliton
transmission matrices without zero components. The other `minimal' solutions
without zeros follow the same pattern but they are not relevant to
the discussion of transmission matrices because they are not
invertible. The notation used for them is abbreviated and omits the
Kronecker deltas.

\begin{itemize}
\item $t_{21}\,t_{13}=t_{23},\quad
\frac{t_{32}\,t_{21}}{t_{22}}=t_{31},\quad
\frac{t_{23}\,t_{31}}{t_{33}}=t_{21},$
\begin{equation}
\left(%
\begin{array}{ccc}\label{IIAIIBIC}
  Q^{\alpha\cdot l_1}
  & t_{12}\,x^{4/3}
  & t_{13}\,x^{2/3}\,Q^{\alpha\cdot l_1} \\
  t_{21}\,x^{2/3}\,Q^{-\alpha\cdot l_3}
  & t_{22}\,Q^{\alpha\cdot l_2}
  & t_{23}\,x^{4/3}\,Q^{-\alpha\cdot l_3} \\
 t_{31}\,x^{4/3}\,Q^{-\alpha\cdot l_3}
  & t_{32}\,x^{2/3}\,Q^{\alpha\cdot l_2}
  & t_{33}\,x^{2}\,Q^{-\alpha\cdot l_3} \\
\end{array}%
\right)
\end{equation}

\item $t_{12}\,t_{31}=t_{32},\,\,
\frac{t_{23}\,t_{12}}{t_{22}}=t_{13},\,\,
\frac{t_{32}\,t_{13}}{t_{33}}=t_{12},$
\begin{equation}
\left(%
\begin{array}{ccc}
 Q^{-\alpha\cdot l_1}
  & t_{12}\,x^{-2/3}\,Q^{\alpha\cdot l_3}
  & t_{13}\,x^{-4/3}\,Q^{\alpha\cdot l_3} \\
  t_{21}\,x^{-4/3}
  & t_{22}\,Q^{-\alpha\cdot l_2}
  & t_{23}\,x^{-2/3}\,Q^{-\alpha\cdot l_2} \\
   t_{31}\,x^{-2/3}\,Q^{-\alpha\cdot l_1}
  & t_{32}\,x^{-4/3}\,Q^{\alpha\cdot l_3}
  & t_{33}\,x^{-2}\,Q^{\alpha\cdot l_3} \\
\end{array}%
\right)
\end{equation}

\item $t_{21}\,t_{13}=t_{23},\,\,
\frac{t_{23}\,t_{12}}{t_{22}}=t_{13},\,\,
\frac{t_{32}\,t_{13}}{t_{33}}=t_{12},$
\begin{equation}
\left(%
\begin{array}{ccc}
 Q^{\alpha\cdot l_1}
  & t_{12}\,x^{4/3}\,Q^{-\alpha\cdot l_2}
  & t_{13}\,x^{2/3}\,Q^{-\alpha\cdot l_2} \\
  t_{21}\,x^{2/3}\,Q^{\alpha\cdot l_1}
  & t_{22}\,x^{2}\,Q^{-\alpha\cdot l_2}
  & t_{23}\,x^{4/3}\,Q^{-\alpha\cdot l_2} \\
t_{31}\,x^{4/3}
  & t_{32}\,x^{2/3}\,Q^{\alpha\cdot l_3}
  & t_{33}\,Q^{\alpha\cdot l_3} \\
\end{array}%
\right)
\end{equation}

\item $t_{12}\,t_{31}=t_{32},\,\,
\frac{t_{32}\,t_{21}}{t_{22}}=t_{31},\,\,
\frac{t_{23}\,t_{31}}{t_{33}}=t_{21},$
\begin{equation}
\left(%
\begin{array}{ccc}
 Q^{-\alpha\cdot l_1}
  & t_{12}\,x^{-2/3}\,Q^{-\alpha\cdot l_1}
  & t_{13}\,x^{-4/3} \\
  t_{21}\,x^{-4/3}\,Q^{\alpha\cdot l_2}
  & t_{22}\,x^{-2}\,Q^{\alpha\cdot l_2}
  & t_{23}\,x^{-2/3}\,Q^{-\alpha\cdot l_3} \\
 t_{31}\,x^{-2/3}\,Q^{\alpha\cdot l_2}
  & t_{32}\,x^{-4/3}\,Q^{\alpha\cdot l_2}
  & t_{33}\,Q^{-\alpha\cdot l_3} \\
\end{array}%
\right)
\end{equation}
 \item $t_{21}\,t_{13}=t_{23},\,\,
\frac{t_{23}\,t_{12}}{t_{22}}=t_{13},\,\,
\frac{t_{23}\,t_{31}}{t_{33}}=t_{21},$
\begin{equation}
\left(%
\begin{array}{ccc}
 Q^{\alpha\cdot l_1}
 & t_{12}\,x^{4/3}\,Q^{-\alpha\cdot l_2}
 & t_{13}\,x^{2/3}\,Q^{\alpha\cdot l_1} \\
 t_{21}\,x^{2/3}\,Q^{\alpha\cdot l_1}
 & t_{22}\,x^{2}\,Q^{-\alpha\cdot l_2}
 & t_{23}\,x^{4/3}\,Q^{\alpha\cdot l_1} \\
t_{31}\,x^{4/3}\,Q^{-\alpha\cdot l_3}
 & t_{32}\,x^{2/3}
 & t_{33}\,x^{2}\,Q^{-\alpha\cdot l_3} \\
\end{array}%
\right)
\end{equation}

\item $t_{12}\,t_{31}=t_{32},\,\,
\frac{t_{32}\,t_{21}}{t_{22}}=t_{31},\,\,
\frac{t_{32}\,t_{13}}{t_{33}}=t_{12},$
\begin{equation}
\left(%
\begin{array}{ccc}
Q^{-\alpha\cdot l_1}
 & t_{12}\,x^{-2/3}\,Q^{-\alpha\cdot l_1}
 & t_{13}\,x^{-4/3}\,Q^{\alpha\cdot l_3} \\
 t_{21}\,x^{-4/3}\,Q^{\alpha\cdot l_2}
 & t_{22}\,x^{-2}\,Q^{\alpha\cdot l_2}
 & t_{23}\,x^{-2/3} \\
t_{31}\,x^{-2/3}\,Q^{-\alpha\cdot l_1}
 & t_{32}\,x^{-4/3}\,Q^{-\alpha\cdot l_1}
 & t_{33}\,x^{-2}\,Q^{\alpha\cdot l_3} \\
\end{array}%
\right).
\end{equation}
\end{itemize}

\p It is not difficult to check the invertibility for
these matrices. Consider an infinite-dimensional matrix ${\cal A}$  of the general
type under consideration:

\begin{equation}
{\cal A}=\left(%
\begin{array}{lll}
a_{11}(\alpha)\delta_{\alpha}^{\beta}
  & a_{12}(\alpha)\delta_{\alpha}^{\beta-\alpha_1}
  & a_{13}(\alpha) \delta_{\alpha}^{\beta+\alpha_0}\phantom{} \\
  a_{21}(\alpha)\delta_{\alpha}^{\beta+\alpha_1}\phantom{}
  &a_{22}(\alpha)\delta_{\alpha}^{\beta}\phantom{}
  & a_{23}(\alpha)\delta_{\alpha}^{\beta-\alpha_2}\phantom{} \\
 a_{31}(\alpha)\delta_{\alpha}^{\beta-\alpha_0}\phantom{}
  & a_{32}(\alpha)\delta_{\alpha}^{\beta+\alpha_2}\phantom{}
  & a_{33}(\alpha)\delta_{\alpha}^{\beta}\phantom{} \\
\end{array}
\right)\nn
\end{equation}
this is invertible if and only if, for every $\alpha$,
\begin{eqnarray}
&&a_{11}(\alpha)\left[a_{22}(\alpha+\alpha_1)a_{33}(\alpha+\alpha_1+\alpha_2)-
a_{23}(\alpha+\alpha_1)a_{32}(\alpha+\alpha_1+\alpha_2)\right]\nn \\
&&\ \ - a_{12}(\alpha)\left[a_{21}(\alpha+\alpha_1)a_{33}(\alpha+\alpha_1+\alpha_2)-
a_{23}(\alpha+\alpha_1)a_{31}(\alpha+\alpha_1+\alpha_2)\right]\nn \\
&&\ \ \ \ \ \  +a_{31}(\alpha)\left[a_{21}(\alpha+\alpha_1)a_{32}(\alpha+\alpha_1+\alpha_2)-
a_{22}(\alpha+\alpha_1)a_{31}(\alpha+\alpha_1+\alpha_2)\right]\ne 0.\nn
\end{eqnarray}
This is similar to the determinant condition for a finite dimensional matrix,
but note the shifts in the arguments of the elements.
Using this condition, it is easy to demonstrate that only the first two solutions listed
above are invertible.

\p The full family of solutions of which \eqref{IIAIIBIIC} is the
`minimal' example is given by
\begin{itemize}
\item $t_{21}\,t_{13}\,Q^{\alpha_1 \cdot \hat{l}_2}=t_{23},\qquad
\frac{t_{32}\,t_{21}}{t_{22}}\,Q^{\alpha_2 \cdot
\hat{l}_3}=t_{31},\qquad \frac{t_{32}\,t_{13}}{t_{33}}\,Q^{\alpha_2
\cdot \hat{l}_2}=t_{12},$

\begin{equation}\label{IIAIIBIICfamily}
\left(%
\begin{array}{ccc}
Q^{\alpha \cdot l_1}\,\delta_{\alpha}^{\beta}\phantom{}
  \,
  & t_{12}\,Q^{\alpha \cdot \hat{l}_3}\,x^{4/3}\,\delta_{\alpha}^{\beta-\alpha_1}\,\phantom{}
  & t_{13}\,x^{2/3}\,Q^{-\alpha \cdot (l_2+\hat{l}_2)}
  \, \delta_{\alpha}^{\beta+\alpha_0}\phantom{} \\
  t_{21}\,x^{2/3}\,Q^{-\alpha \cdot (l_3+\hat{l}_3)}
  \,\delta_{\alpha}^{\beta+\alpha_1}\phantom{}
  &
  t_{22}\,Q^{\alpha \cdot l_2}\,\delta_{\alpha}^{\beta}\phantom{}
  & t_{23}\,Q^{\alpha \cdot \hat{l}_1}\,x^{4/3}\,\delta_{\alpha}^{\beta-\alpha_2}\phantom{} \\
 t_{31}\,Q^{\alpha \cdot \hat{l}_2}\,x^{4/3}\,\delta_{\alpha}^{\beta-\alpha_0}\phantom{}
  & t_{32}\,x^{2/3}\,Q^{-\alpha\cdot (l_1+\hat{l}_1)}
  \,\delta_{\alpha}^{\beta+\alpha_2}\phantom{}
  & t_{33}\,Q^{\alpha\cdot l_3}
  \,\delta_{\alpha}^{\beta}\phantom{} \\
\end{array}
\right)
\end{equation}

\end{itemize}
where $\hat{l}_{p}=(\hat{m}_p\,\alpha_1+\hat{n}_p\,\alpha_2)$ are
vectors lying in the weight lattice, such that
\begin{equation}\label{extraweightconstraints} \alpha_1\cdot
\hat{l}_{1}=\alpha_2\cdot \hat{l}_{3},\qquad\alpha_2\cdot
\hat{l}_{3}=\alpha_0\cdot \hat{l}_{1},\qquad\alpha_0\cdot
\hat{l}_{3}=\alpha_1\cdot \hat{l}_{2}.
\end{equation}
It should be noticed that the extra dependence of $\alpha$ in the off-diagonal
entries of the matrix \eqref{IIAIIBIICfamily} as compared with corresponding
elements in \eqref{IIAIIBIIC} does not affect the constraints coming from
 groups 4 and 5 (dealing with the $\alpha$ dependence), due to
 \eqref{extraweightconstraints}. Setting
$\hat{l}_{1}=\hat{l}_{2}=\hat{l}_{3}=0$, the solution
\eqref{IIAIIBIIC} is recovered. It is in this sense that \eqref{IIAIIBIIC} is
considered a `minimal' solution.

\p
Finally, it should be pointed out that each solution found can be
multiplied by an overall function of $\theta$ that cannot be
determined by the triangle equations alone.

\p In addition to all the above solutions, there are others that
 allow some entries of the $T$ matrix to be set equal to zero.
Suppose  $K=I=W=0$, or $J=L=V=0$, then the previous
analysis has to be modified. In this  situation, the relations
in group 2 state that the following ratios
\begin{equation}
\frac{a_{\alpha}}{b_{\alpha}}, \qquad \frac{a_{\alpha}}{c_{\alpha}},
\qquad \frac{b_{\alpha}}{c_{\alpha}},\nn
\end{equation}
are independent of rapidity. Concerning  group 3, because of the
presence of $T$ matrix elements equal to zero, not all constraints
listed in \eqref{constraint3A1}-\eqref{constraint3C2} survive.
However, the ones which do are unmodified. Equations in group 4
force relations \eqref{constraint4} to hold, as before. In
addition, if $K=I=W=0$, they imply the ratios
\begin{equation}
\frac{a_{\alpha}}{l_{\alpha}}x^{2/3}, \qquad
\frac{b_{\alpha}}{v_{\alpha}}x^{2/3},\qquad
\frac{c_{\alpha}}{j_{\alpha}}x^{2/3},
\end{equation}
are independent of rapidity, and  only one combination
of relations in the subgroups $3A, 3B, 3C$ is allowed, namely
\begin{equation}
\frac{a_{\alpha+\alpha_1}}{a_{\alpha}}=\frac{1}{q},\quad
\frac{a_{\alpha+\alpha_0}}{a_{\alpha}}=q,\quad
\frac{b_{\alpha+\alpha_1}}{b_{\alpha}}=q,\quad
\frac{b_{\alpha+\alpha_0}}{b_{\alpha}}=\frac{1}{q},\quad
\frac{c_{\alpha+\alpha_0}}{c_{\alpha}}=\frac{1}{q},\quad
\frac{c_{\alpha+\alpha_2}}{c_{\alpha}}=q,\nn
\end{equation}
with
\begin{equation}
\frac{h_v(\alpha_1)}{h_j(\alpha_0)}=
\frac{h_l(\alpha_1)}{h_j(\alpha_2)}=
\frac{h_v(\alpha_2)}{h_l(\alpha_0)}=1.\nn
\end{equation}
On the other hand, if $J=L=V=0$, the ratios
\begin{equation}
\frac{a_{\alpha}}{i_{\alpha}}x^{-2/3}, \qquad
\frac{b_{\alpha}}{w_{\alpha}}x^{-2/3},\qquad
\frac{c_{\alpha}}{k_{\alpha}}x^{-2/3},
\end{equation}
are independent of rapidity and only the following combinations of
relations in the subgroups $3A, 3B, 3C$ are permitted, namely
\begin{equation}
\frac{a_{\alpha+\alpha_1}}{a_{\alpha}}=q,\quad
\frac{a_{\alpha+\alpha_0}}{a_{\alpha}}=\frac{1}{q},\quad
\frac{b_{\alpha+\alpha_1}}{b_{\alpha}}=\frac{1}{q},\quad
\frac{b_{\alpha+\alpha_0}}{b_{\alpha}}=q,\quad
\frac{c_{\alpha+\alpha_0}}{c_{\alpha}}=q,\quad
\frac{c_{\alpha+\alpha_2}}{c_{\alpha}}=\frac{1}{q},\nn
\end{equation}
with
\begin{equation}
\frac{h_w(\alpha_1)}{h_k(\alpha_0)}=
\frac{h_i(\alpha_1)}{h_k(\alpha_2)}=
\frac{h_i(\alpha_0)}{h_w(\alpha_2)}=1.\nn
\end{equation}
Finally, the relations in group 5 disappear completely.

\p The two `minimal' solutions of this type are
\begin{equation}
\left(%
\begin{array}{ccc}\label{KIW}
  Q^{\alpha\cdot l_1}\,\delta_{\alpha}^{\beta}\phantom{}
  & 0\phantom{aaaa}
  & t_{13}\,x^{2/3}\,
  \delta_{\alpha}^{\beta+\alpha_0} \\
  t_{21}\,x^{2/3}\,
  \delta_{\alpha}^{\beta+\alpha_1}
  & t_{22}\,Q^{\alpha\cdot l_2}\,
  \delta_{\alpha}^{\beta}\phantom{}
  & 0\phantom{aaaa} \\
 0\phantom{aaaa}
  & t_{32}\,x^{2/3}\,
  \delta_{\alpha}^{\beta+\alpha_2}
  & t_{33}\,Q^{\alpha\cdot l_3}\,
  \delta_{\alpha}^{\beta}\phantom{} \\
\end{array}%
\right)
\end{equation}
and
\begin{equation}\label{VJL}
\left(%
\begin{array}{ccc}
  Q^{-\alpha\cdot l_1}\,\delta_{\alpha}^{\beta}\phantom{}
  & t_{12}\,x^{-2/3}\,
  \delta_{\alpha}^{\beta-\alpha_1}
  & 0\phantom{aaaa} \\
  0\phantom{aaaa}
  & t_{22}\,Q^{-\alpha\cdot l_2}\,
  \delta_{\alpha}^{\beta}\phantom{}
  & t_{23}\,x^{-2/3}\,
  \delta_{\alpha}^{\beta-\alpha_2} \\
 t_{31}\,x^{-2/3}\,
 \delta_{\alpha}^{\beta-\alpha_0}
  & 0\phantom{aaaa}
  & t_{33}\,Q^{-\alpha\cdot l_3}\,
  \delta_{\alpha}^{\beta}\phantom{} \\
\end{array}%
\right)
\end{equation}
and these are of relevance to the defect for the reasons explained earlier.

\p For these types of solution there are also general families.
 For example, \eqref{VJL} belongs to
the following set
\begin{equation}
\left(%
\begin{array}{ccc}
  Q^{\alpha\cdot l_1}\,\delta_{\alpha}^{\beta}\phantom{}
  & 0\phantom{aaaa}
  & t_{13}\,x^{2/3}\,Q^{-\alpha\cdot \hat{l}_{2}}\,
  \delta_{\alpha}^{\beta+\alpha_0} \\
  t_{21}\,x^{2/3}\,Q^{-\alpha\cdot \hat{l}_{3}}\,
  \delta_{\alpha}^{\beta+\alpha_1}
  & t_{22}\,Q^{\alpha\cdot l_2}\,
  \delta_{\alpha}^{\beta}\phantom{}
  & 0\phantom{aaaa} \\
 0\phantom{aaaa}
  & t_{32}\,x^{2/3}\,Q^{-\alpha\cdot \hat{l}_{1}}\,
  \delta_{\alpha}^{\beta+\alpha_2}
  & t_{33}\,Q^{\alpha\cdot l_3}\,
  \delta_{\alpha}^{\beta}\phantom{} \\
\end{array}%
\right)
\end{equation}
where the vectors $\hat{l}_{p}$ satisfy
\eqref{extraweightconstraints}.

\end{document}